\documentclass[journal,onecolumn]{IEEEtran}

\usepackage{amsthm}
\usepackage{subfigure}
\usepackage{slashbox}
\usepackage{graphicx}
\usepackage{amssymb,amstext,amsmath}
\usepackage{float}
\usepackage{bbm}
\usepackage{bm}
\usepackage{multirow}
\usepackage{makecell}
\usepackage{tabu}
\usepackage{color}
\usepackage{caption}
\usepackage{verbatim}
\usepackage{enumitem}
\usepackage{tikz}
\usetikzlibrary{positioning,arrows}
\tikzset{
block/.style={
  draw,
  rectangle,
  minimum height=0.8cm,
  minimum width=0.8cm, align=center
  },
line/.style={->,>=latex'}
}
\usetikzlibrary{arrows}

\newcommand{\beq}{\begin{equation}}
\newcommand{\eeq}{\end{equation}}
\newcommand{\beqa}{\begin{eqnarray}}
\newcommand{\eeqa}{\end{eqnarray}}

\newcommand{\bxi}{{\mbox{\boldmath $\bxi$}}}

\newcommand{\dfn}{\triangleq}

\newcommand{\rw}{\rightarrow}

\newcommand{\mbbQ}{\mathbb{Q}}
\newcommand{\Real}{\mathbb{R}}

\newcommand{\mA}{{\mathcal A}}
\newcommand{\mB}{{\mathcal B}}

\newcommand{\mL}{{\mathcal L}}

\newcommand{\mI}{{\mathcal I}}

\newcommand{\mX}{{\mathcal X}}
\newcommand{\mY}{{\mathcal Y}}
\newcommand{\mZ}{{\mathcal Z}}
\newcommand{\mN}{{\mathcal N}}

\newcommand{\bfa}{\textit{\textbf{a}}}
\newcommand{\bfu}{\textit{\textbf{u}}}
\newcommand{\x}{\textit{\textbf{x}}}
\newcommand{\y}{\textit{\textbf{y}}}
\newcommand{\z}{\textit{\textbf{z}}}

\newcommand{\X}{\textit{\textbf{X}}}
\newcommand{\Y}{\textit{\textbf{Y}}}

\newcommand{\mfL}{($\mathfrak{L}$)}
\newcommand{\mfD}{($\mathfrak{D}$)}
\newcommand{\mfC}{($\mathfrak{C}$)}

\newcommand{\sff}{{\sf f}}
\newcommand{\sfF}{{\sf F}}

\newtheorem{Teorema}{\em Theorem}
\newtheorem{Definicion}{\em Definition}
\newtheorem{Lema}{\em Lemma}
\newtheorem{Proposicion}{\em Proposition}

\newtheorem{Nota}{\em Remark}
\newtheorem{Algoritmo}{\em Algorithm}

\begin{document}

\title{Adapting the Number of Particles in Sequential Monte Carlo Methods through an Online Scheme for Convergence Assessment}
 \author{V\'ictor Elvira,~\IEEEmembership{Member,~IEEE}, Joaqu\'in M\'iguez, Petar M. Djuri\'c,~\IEEEmembership{Fellow,~IEEE}
 \thanks{{V. Elvira is with IMT Lille Douai (Institut Mines-T\'el\'ecom) and CRIStAL laboratory (France), e-mail: victor.elvira@imt-lille-douai-fr.} {J. M\'iguez is with the Department of Signal Theory and Communications, Universidad Carlos III de Madrid (Spain), e-mail: joaquin.miguez@uc3m.es.} P. M. Djuri\'c is with the Department of Electrical and Computer Engineering, Stony Brook University (USA), e-mail: petar.djuric@stonybrook.edu. This work was partially supported by {\em Ministerio de Econom\'ia y Competitividad} of Spain (TEC2013-41718-R OTOSiS, TEC2012-38883-C02-01 COMPREHENSION, and TEC2015-69868-C2-1-R ADVENTURE), the Office of Naval Research Global (N62909-15-1-2011), and the National Science Foundation (CCF-1320626 and CCF-1618999).
	}

}

\markboth{IEEE TRANSACTIONS ON SIGNAL PROCESSING}%
{Shell \MakeLowercase{\textit{et al.}}: Bare Demo of IEEEtran.cls for Journals}
\maketitle

\begin{abstract}
Particle filters are broadly used to approximate posterior distributions of hidden states in state-space models by means of sets of weighted particles. While the convergence of the filter is guaranteed when the number of particles tends to infinity, the quality of the approximation is usually unknown but strongly dependent on the number of particles. In this paper, we propose a novel method for assessing the convergence of particle filters in an online manner, as well as a simple scheme for the online adaptation of the number of particles based on the convergence assessment. The method is based on a sequential comparison between the actual observations and their predictive probability distributions approximated by the filter. We provide a rigorous theoretical analysis of the proposed methodology and, as an example of its practical use, we present simulations of a simple algorithm for the dynamic and online adaptation of the number of particles during the operation of a particle filter on a stochastic version of the Lorenz system.
\end{abstract}
\begin{IEEEkeywords}
Particle filtering, sequential Monte Carlo, convergence assessment, predictive distribution, convergence analysis, computational complexity, adaptive complexity.
\end{IEEEkeywords}

\IEEEpeerreviewmaketitle

%%%%%%%%%%%%%%%%%%%%%%
%%%%%%%%%%%%%%%%%%%%%%
%%%%%%%%%%%%%%%%%%%%%%
\section{Introduction}
\label{sec_introduction}
%%%%%%%%%%%%%%%%%%%%%%
%%%%%%%%%%%%%%%%%%%%%%
%%%%%%%%%%%%%%%%%%%%%%

%%%
%
%%%
\subsection{Background}

Many problems in science and engineering can be described by dynamical models where hidden states of the systems change over time and observations that are functions of the states are available.
Often, the observations are sequentially acquired and the interest is in making recursive inference on the hidden states. 
In many applications, the Bayesian approach to the problem is adopted because it allows for optimal inclusion of prior knowledge of the unknown state in the estimation process \cite{West96,Ristic04}. 
In this case, the prior information and the likelihood function that relates the hidden state and the observation are combined yielding a posterior distribution of the state.

Exact Bayesian inference, however, is only possible in a small number of scenarios, including linear Gaussian state-space models (using the Kalman filter \cite{Kalman60,Anderson79}) and finite state-space hidden Markov models (HMM filters \cite{rabiner1986introduction}).
Therefore, in many other practical problems, only approximate inference methods can be used. 
One class of suboptimal methods is particle filtering, which is also known as sequential Monte Carlo sampling \cite{Liu98,Doucet00,Doucet01b,Djuric03,Cappe07}. Since the publication of \cite{Gordon93}, where the sampling importance resampling (SIR) filter was introduced, particle filtering has received outstanding attention in research and practice. 
Particle filters approximate posterior distributions of the hidden states sequentially and recursively. They do it by exploiting the principle of importance sampling and by using sets of weighted particles \cite{Liu98,Doucet00,Bain08}.	 

One key parameter of particle filters is the number of particles. It can be proved that the rate of convergence of the approximate probability distribution towards the true posterior is inversely proportional to the square root of the number of particles used in the filter \cite{Bain08,DelMoral00}.
This, too, entails that the filter ``perfectly'' approximates the posterior distribution when the number of particles tends to infinity.
However, since the computational cost grows with the number of particles, practitioners must choose a specific number of particles in the design of their filters.

In many applications, the observations arrive sequentially, and there is a strict deadline for processing each new observation.
Then, one could argue that the best solution in terms of filter performance is to increase the number of particles as much as possible and keep it fixed.
Also, in some hardware implementations, the number of particles is a design parameter that cannot be modified during implementation.
Nevertheless, in many other applications where resources are scarce or are shared with a dynamical allocation and/or with energy restrictions, one might be interested in adapting the number of particles in a smart way. One would use enough particles to achieve a certain performance requirement but without wasting resources by using many more particles if they do not translate into a significant improvement of the filter performance.

The selection of the number of particles, however, is often a delicate subject because, (1) the performance of the filter (the quality of the approximation) cannot usually be described in advance as a function of the number of particles, and (2) the mismatch between the approximation provided by the filter and the unknown posterior distribution is obviously also unknown.
Therefore, although there is a clear trade-off between performance and computational cost, this relation is not straightforward; e.g., increasing the number of particles over a certain value may not significantly improve the quality of the approximation while decreasing the number of particles below some other value can dramatically affect the performance of the filter.

Few papers in the wide literature have addressed the problem of online assessment of the filter convergence for the purpose of adapting the number of particles. In \cite{fox2003adapting}, the number of particles is selected so that a bound on the  approximation error does not exceed a threshold with certain probability. The latter error is defined as the Kullback-Leibler divergence (KLD) between the approximate filter distribution and a grid-discretized version of the true one (which is itself a potentially-costly approximation with an unknown error).  In \cite{soto2005self}, an adaptation of the number of particles is proposed, based on the KLD approach of \cite{fox2003adapting} and an estimate of the variance of the estimators computed via the particle filter, along with an improvement of the proposal distributions. In \cite{straka2006particle}, the adaptation of the number of particles is based on the effective sample size. These methods are heuristic: they do not enjoy any theoretical guarantees (in the assessment of the approximation errors made by the particle filter) and the allocation of particles, therefore, cannot be ensured to be optimal according to any probabilistic criterion. Some techniques based on more solid theoretical ground have been proposed, within the applied probability community, during  the last few years. We discuss them below.

Two types of unbiased estimators of the variance in the approximation of integrals using a class of particle filters were analyzed in \cite{Lee15} using the Feynman-Kac framework of \cite{DelMoral04}. As an application of these results, it was suggested to use these estimators to select the number of particles in the filter. In particular, the scheme proposed in \cite{Lee15} is a batch procedure in which a particle filter is run several times over the whole data sequence, with increasing number of particles, until the variance of the integral of interest is found to fall below a prescribed threshold. This approach cannot be used for online assessment, which is the goal of the present paper. Another batch method (thus, also not applicable for online assessment) for particle allocation has been recently proposed in \cite{Bhadra16}, where an ad hoc autoregressive model is fitted to estimate the variance of the estimators produced by the particle filter.

Papers on so-called \textit{alive} particle filters can also be found in the literature \cite{LeGland05,Jasra13,DelMoral15alive}. These articles focus on models where the likelihood function can take zero value for some regions of the state space, in such a way that there is the risk that a collection of zero-weight particles are generated if a standard algorithm is employed. To avoid this limitation, alive particle filters are based on sampling schemes where new particles are generated until a prescribed number of them, $M$, attain non-zero weights. The computational cost of the algorithm per time step is, therefore, random. Moreover, the number $M$ is chosen a priori and there is no assessment of whether $M$ allows for reaching adequate accuracy of the estimators (the methodology proposed in the present manuscript can be directly applied to alive particle filters in order to adapt $M$).

In order to guarantee that the particle set yields a sufficiently good representation, in \cite{Hu08} it is proposed to test whether the particle estimate of the predictive density of the observation at time $t$ given the previous data is sufficiently large, i.e., whether it is above a prescribed (heuristically chosen) threshold. When the particle set does not satisfy this condition, it is discarded and a new collection of particles is generated. The number of particles is not adapted, since all generated sets have the same size. The computational cost of this algorithm is random.

Finally, in \cite[Chapter 4]{Cornebise09thesis} it is proposed to use the coefficient of variation of the weights (or, equivalently, the effective sample size) in order to detect those observations for which there is a large $\chi^2$-divergence between the proposal distribution used to generate the set of particles and the target distribution. This connection is rigorously established in \cite{Cornebise09thesis}. The algorithm, however, is computationally costly compared to classical methods: at each time step, a complete set of particles and weights are generated, and the coefficient of variation is computed. If this coefficient is too high, the particles are discarded, the algorithm ``rolls back,'' and a new, larger set of particles is generated for better representation of the target distribution (this step is termed ``refuelling'' in \cite{Cornebise09thesis}). Although the algorithm enjoys theoretical guarantees, it relies on keeping the particle approximation ``locked'' to the target distribution at all times. It is known that, once the particle filter has lost track of the state distribution, the effective sample size (and, hence, coefficient of variation) becomes uninformative \cite{Beskos14stability} and, therefore, the link with the $\chi^2$-divergence is lost.

%%%
%
%%%
\subsection{Contributions}

We introduce a model--independent methodology for the online assessment of the convergence of particle filters and carry out a rigorous analysis that ensures the consistency of the proposed scheme under fairly standard assumptions. The method is an extension of our previous work presented in \cite{djuric2010assessment}. In the proposed scheme, the observations are processed one at a time and the filter performance is assessed by measuring the discrepancy between the actual observation at each time step and a number of fictitious data-points drawn from the particle approximation of the predictive probability distribution of the observations.
The method can be exploited to adjust the number of particles dynamically when the performance of the filter degrades below a certain desired level. This would allow a practitioner to select the operation point by considering performance-computational cost tradeoffs.
Based on the method, we propose a simple and efficient algorithm that adjusts the number of particles in real time. We demonstrate the performance of the algorithm numerically by running it for a stochastic version of the $3$-dimensional Lorenz 63 system. As already noted, this paper builds on the method from \cite{djuric2010assessment}. However, the main difference here is that the underlying model is not questioned -- instead, it is {\em assumed} to be correct. The connection between \cite{djuric2010assessment} and the present work is that they both build upon the ability to compute predictive statistics of the upcoming observations that turn out to be independent of the underlying state space model. In this paper we have rigorous theoretical results regarding the particle approximations of the predictive distribution of the observations (while this issue was ignored in \cite{djuric2010assessment}). Finally, we suggest practical schemes for the online adjustment of the number of particles.

Let us point out that the adaptive procedure for the online selection of the number of particles described herein is only one of many that can exploit the results of the convergence analysis. In other words, our analysis opens the door for development of new families of algorithms for online adaptation of the number of particles by way of online convergence assessment.

%%%
%
%%%
\subsection{Organization of the paper}

The rest of the paper is organized as follows. In Section \ref{sPF} we describe the class of state space Markov models and provide a basic  background on the well-known bootstrap particle filter of \cite{Gordon93}. The theoretical results that enable the online assessment of particle filters are stated in Section \ref{sec:convergence}, with full details and proofs contained in Appendix \ref{apTh1}. The proposed methodology for online convergence assessment of the particle filter is introduced in Section \ref{sec_method}. Furthermore, this section provides a simple algorithm for the dynamic, online adaptation of  the number of particles. In Section \ref{section_numerical}, we illustrate the validity of the method by means of computer simulations for a stochastic Lorenz 63 model. Finally, Section \ref{sec_conclusions} contains a summary of results and some concluding remarks.

%%%%%%%%%%%%%%%%%%%%%%
%%%%%%%%%%%%%%%%%%%%%%
%%%%%%%%%%%%%%%%%%%%%%
\section{Particle filtering}
\label{sPF}
%%%%%%%%%%%%%%%%%%%%%%
%%%%%%%%%%%%%%%%%%%%%%
%%%%%%%%%%%%%%%%%%%%%%

In this section we describe the class of state space models of interest and then present the standard particle filter (PF), which is the basic building block for the methods to be introduced later. % and introduce a new asymptotic convergence result that is key to the design and analysis of the proposed online convergence assessment scheme introduced in Section \ref{sec_method}.

%%%%
\subsection{State space models and stochastic filtering} \label{ssSSPF}
%%%%

Let us consider discrete-time, Markov dynamic systems in state-space form described by the triplet\footnote{In most of the paper we abide by a simplified notation where $p(x)$ denotes the probability density function (pdf) of the random variable $X$. This notation is argument-wise, hence if we have two random variables $X$ and $Y$, then $p(x)$ and $p(y)$ denote the corresponding density functions, possibly different; $p(x,y)$ denotes the joint pdf and $p(x|y)$ is the conditional pdf of $X$ given $Y=y$. A more accurate notation, which avoids ambiguities, is used for the analysis and the statement of the theoretical results. Besides, vectors are denoted by bold-face letters, e.g., $\x$, while regular-face is used for scalars, e.g., $x$.}
\begin{eqnarray}
\X_0 &\sim& p(\x_0), \label{eqPrior}\\
\X_t &\sim& p(\x_t|\x_{t-1}), \label{eqState}\\
\Y_t &\sim& p(\y_t|\x_t), \label{eqLikelihood}
\end{eqnarray}
where 
\begin{itemize}
\item $t \in \mathbb{N}$ denotes discrete time; 
\item $\X_t$ is the $d_x \times 1$-dimensional (random) system state at time $t$, which takes variables in the set $\mX \subseteq \Real^{d_x}$,
\item $p(\x_0)$ is the a priori pdf of the state, while
\item $p(\x_t|\x_{t-1})$ denotes the conditional density of the state $\X_t$ given $\X_{t-1}=\x_{t-1}$;
\item $\Y_t$ is the $d_y \times 1$-dimensional observation vector at time $t$, which takes values in the set $\mY \subseteq \Real^{d_y}$ and is assumed to be conditionally independent of all other observations given the state $\X_t$, 
\item $p(\y_t|\x_t)$ is the conditional pdf of $\Y_t$ given $\X_t=\x_t$. It is often referred to as the {\em likelihood} of $\x_t$, when it is viewed as a function of $\x_t$ given $\y_t$.
\end{itemize}
The model described by Eqs. \eqref{eqPrior}--\eqref{eqLikelihood} includes a broad class of systems, both linear and nonlinear, with Gaussian or non-Gaussian perturbations. Here we focus on the case where all the model parameters are known. However, the proposed method can also be used for models with unknown parameters for which suitable particle filtering methods are available \cite{Chen00,Chopin12,Crisan13b}. We assume that the prior distribution of the state $p(\x_0)$ is also known.

The stochastic filtering problem consists in the computation of the sequence of posterior probability distributions given by the so-called {\em filtering} densities $p(\x_t|\y_{1:t})$, $t=1, 2, \cdots$. The pdf $p(\x_t|\y_{1:t})$ is closely related to the one-step-ahead predictive state density $p(\x_t|\y_{1:t-1})$, which is of major interest in many applications and can be written down by way of the Chapman-Kolmogorov equation,
\begin{equation}
p(\x_t|\y_{1:t-1}) = \int p(\x_t|\x_{t-1}) p(\x_{t-1}|\y_{1:t-1}) d\x_{t-1}. \label{eqCK}
\end{equation}
Using Bayes' theorem together with Eq. \eqref{eqCK}, we obtain the well-known recursive factorization of the filtering pdf
\begin{equation}
p(\x_t|\y_{1:t}) \propto p(\y_t|\x_t) \int p(\x_t|\x_{t-1}) p(\x_{t-1}|\y_{1:t-1}) d\x_{t-1}.
\nonumber
\end{equation}
For conciseness and notational accuracy, we use the measure-theoretic notation 
\begin{equation}
\pi_t(d\x_t):=p(\x_t|\y_{1:t})d\x_t, \quad \xi_t(d\x_t):=p(\x_t|\y_{1:t-1})d\x_t
\nonumber
\end{equation}
to represent the filtering and the predictive posterior probability distributions of the state, respectively. Note that $\pi_t$ and $\xi_t$ are probability measures, hence, given a Borel set $A \subset \mX$, $\pi_t(A)=\int_A \pi(d\x_t)$ and $\xi_t(A) = \int_A \xi_t(d\x_t)$ denote the posterior probability of the event $\X_t \in A$ conditional on $\Y_{1:t}=\y_{1:t}$ and $\Y_{1:t-1}=\y_{1:t-1}$, respectively.

However, the object of main interest for the convergence assessment method to be introduced in this paper is the predictive pdf of the observations, namely the function $p(\y_t|\y_{1:t-1})$ and the associated probability measure
\begin{equation}
\mu_t(d\y_t) := p(\y_t|\y_{1:t-1}) d\y_t.
\nonumber
\end{equation}
The density $p(\y_t|\y_{1:t-1})$ is the normalization constant of the filtering density $p(\x_t|\y_{1:t})$, and it is related to the predictive state pdf $p(\x_t|\y_{1:t-1})$ through the integral
\begin{equation}
p(\y_t|\y_{1:t-1}) = \int p(\y_t|\x_t) p(\x_t|\y_{1:t-1}) d\x_t. 
\label{eqPy}
\end{equation}
It also plays a key role in model assessment \cite{djuric2010assessment} and model inference  problems \cite{Chopin12,Crisan13b,Andrieu10}, \cite{Koblents15}.

%%%%%%%%%%%%%%%%%%%%%%
\subsection{The standard particle filter}
\label{sec_back}
%%%%%%%%%%%%%%%%%%%%%%

A PF is an algorithm that processes the observations $\{\y_t\}_{t\ge 1}$ sequentially in order to compute Monte Carlo approximations of the sequence of probability measures $\{\pi_t\}_{t\ge 1}$. The simplest algorithm is the so-called {\em bootstrap particle filter} (BPF)  \cite{Gordon93} (see also \cite{Doucet01}), which consists of a recursive importance sampling procedure and a resampling step. The term ``particle'' refers to a Monte Carlo sample in the state space $\mX$, which is assigned an importance weight. Below, we outline the BPF algorithm with $M$ particles.
%%% 
\begin{Algoritmo} \label{alBF}
Bootstrap particle filter.
\begin{enumerate}
\item {\sf Initialization.} At time $t=0$, draw $M$ i.i.d. samples,  $\x_0^{(m)}$, $m=1,\ldots,M$, from the prior $p(\x_0)$.

\item {\sf Recursive step.} Let $\{ \x_{t-1}^{(m)} \}_{m=1}^M$ be the particles at time $t-1$. At time $t$, proceed with the two steps below.
        \begin{enumerate}
        \item For~$m=1, ..., M$,~draw~$\bar \x_t^{(m)}$~from~the model transition pdf $p(\x_t|\x_{t-1}^{(m)})$. Then compute the normalized importance weights
        \begin{equation}
        w_t^{(m)} = \frac{
                p(y_{t}|\bar \x_t^{(m)})
        }{
                \sum_{k=1}^M p(y_{t}|\bar \x_t^{(k)})
        }, \quad m=1, ..., M.
        \end{equation}

        \item Resample $M$ times with replacement: for $m=1,...,M$, let $\x_t^{(m)}= \bar{\x}_t^{(k)}$ with probability $w_t^{(k)}$, where $k \in \{1,...,M\}$.
        \end{enumerate}
\end{enumerate}
\end{Algoritmo}

For the sake of simplicity, in step 2.(b) above we assume that multinomial resampling \cite{Doucet00} is carried out for every $t\ge 1$. The results and methods to be presented in subsequent sections remain valid when resampling is carried out periodically and/or using alternative schemes such as residual \cite{Liu98}, stratified \cite{Carpenter99} or minimum-variance \cite{Crisan01} resampling (see also \cite{Li15}).  

The simple BPF yields several useful approximations. After sampling at step 2.(a), the predictive state probability measure $\xi_t$ can be approximated as
\begin{equation}
\xi_t^M(d\x_t) = \frac{1}{M} \sum_{m=1}^M \delta_{\bar \x_t^{(m)}}(d\x_t),
\nonumber
\end{equation}
where $\delta_{\x}$ denotes the Dirac delta measure located at $\x \in \mX$. The filter measure $\pi_t$ can be similarly approximated, either using the particles and weights computed at step 2.(a) or the resampled particles after step 2.(b), i.e., 
\begin{equation}
\bar \pi_t^M = \sum_{m=1}^M w_t^{(m)} \delta_{\bar \x_t^{(m)}} \quad \mbox{and} \quad
\pi_t^M = \frac{1}{M} \sum_{m=1}^M \delta_{\x_t^{(m)}},
\nonumber
\end{equation}
respectively. In addition, the BPF 
yields natural approximations of the predictive pdf's of $\X_t$ and $\Y_t$ given the earlier observations $\Y_{1:t-1}=\y_{1:t-1}$. If we specifically denote these functions as $\tilde p_t(\x_t):$ $=$ $p(\x_t|\y_{1:t-1})$ and $p_t(\y_t):=$ $p(\y_t|\y_{1:t-1})$,
then we readily obtain their respective estimates as mixture distributions with $M$ mixands, or,  
\begin{eqnarray}
\tilde p_t^M(\x_t) &:=& \sum_{m=1}^M w_{t-1}^M p(\x_t|\x_{t-1}^{(m)}), \quad \mbox{and} \nonumber\\
p_t^M(\y_t) &:=& \frac{1}{M} \sum_{m=1}^M p(\y_t|\bar \x_t^{(m)}), \nonumber
\end{eqnarray}
for any $\x_t \in \mX$ and $\y_t \in \mY$.

%%%%%%%%%%%%%%%%%%%%%%
%%%%%%%%%%%%%%%%%%%%%%
%%%%%%%%%%%%%%%%%%%%%%
\section{A novel asymptotic convergence result}
\label{sec:convergence}
%%%%%%%%%%%%%%%%%%%%%%
%%%%%%%%%%%%%%%%%%%%%%
%%%%%%%%%%%%%%%%%%%%%%

The convergence of the approximate measures, e.g., $\xi_t^M$, towards the true ones is usually assessed in terms of the estimates of 1-dimensional statistics of the corresponding probability distribution. To be specific, let $f:\mX\rw\Real$ be a real integrable function in the state space and denote\footnote{
Let  $(\mZ,\mB(\mZ))$ be a measurable space, where $\mZ \subseteq \Real^d$ for some integer $d \ge 1$ and $\mB(\mZ)$ is the Borel $\sigma$-algebra of subsets of $\mZ$. If $\alpha$ is a measure on $\mB(\mZ)$ and the function $h : \mZ \rw \Real$ is integrable with respect to (w.r.t.) $\alpha$, then we use the shorthand notation $(f,\alpha) := \int f(z) \alpha(dz)$.
}
\begin{equation}
(f,\xi_t) := \int f(\x_t)\xi_t(d\x_t).
\nonumber
\end{equation}
Under mild assumptions on the state space model, it can be proved that 
\begin{equation}
\lim_{M\rw\infty} (f,\xi_t^M) = \lim_{M\rw\infty} \frac{1}{M}\sum_{m=1}^M f(\x_t^{(m)}) = (f,\xi_t)
\label{eqXiM}
\end{equation}
almost surely (a.s.) \cite{DelMoral04,Bain08}.

According to \eqref{eqPy}, the predictive observation pdf $p_t(\y_t)$ is an integral w.r.t. $\xi_t$ and, as a consequence, Eq. \eqref{eqXiM} implies that $\lim_{M\rw\infty} p_t^M(\y) = p_t(\y)$ a.s. and point-wise for every $\y \in \mY$ under mild assumptions \cite{DelMoral04}. However, existing theoretical results {\em do not} ensure that $p_t^M(\y)$ can converge {\em uniformly} on $\mathcal{Y}$ towards $p_t(\y)$ and this fact prevents us from claiming that $\lim_{M\rw\infty} \int h(\y)p_t^M(\y)d\y = \int h(\y)p_t(\y)d\y = (h,\mu_t)$ in some proper sense for integrable real functions $h(\y)$.

Important contributions of this paper are (a) the proof of a.s. convergence of the random probability measure 
\begin{equation}
\mu_t^M(d\y) := p_t^M(\y)d\y
\nonumber
\end{equation}
towards $\mu_t$ (as $M\rw\infty$) under mild regularity assumptions on the state space model, and (b) the provision of explicit error rates. We point out that $\mu_t^M$ is not a classical point-mass Monte Carlo approximation of $\mu_t$ (as, for example, $\pi_t^M$ is an approximation of $\pi_t$). Instead, the measure $\mu_t^M$ is absolutely continuous with respect to the Lebesgue measure (the same as $\mu_t$ itself). If a different reference measure were used to define the pdf's $p(\x_t|\x_{t-1})$ and $p(\y_t|\x_t)$, say $\nu$, then both $\mu_t$ and $\mu_t^M$ would be absolutely continuous with respect to $\nu$. In order to describe how $\mu_t^M$ converges to $\mu_t$ in a rigorous manner , we need to introduce some notation:
%
%The first contribution of this paper is to prove that, under mild regularity assumptions on the state space model, the continuous random probability measure 
%\begin{equation}
%\mu_t^M(d\y) := p_t^M(\y)d\y
%\nonumber
%\end{equation}
%converges a.s. to $\mu_t$ and provide explicit error rates. %To express this result rigorously, we need to introduce some notation:
%
%\cblack{Let us point out that $\mu_t^M$ is not a classical point-mass Monte Carlo approximation of $\mu_t$ (as, for example, $\pi_t^M$ is an approximation of $\pi_t$). Instead, the measure $\mu_t^M$ is absolutely continuous with respect to the Lebesgue measure for the state space model considered in this paper (the same as $\mu_t$ itself). If a different reference measure were used to define the pdf's $p(\x_t|\x_{t-1})$ and $p(\y_t|\x_t)$, say $\nu$, then both $\mu_t$ and $\mu_t^M$ would be absolutely continuous with respect to $\nu$. In order to describe in a rigorous manner how $\mu_t^M$ converges to $\mu_t$, we need to introduce some notation:}
\begin{itemize}
\item For each $t\ge 1$, let us define the function $g_t(\y_t,\x_t) := p(\y_t|\x_t)$, i.e., the conditional pdf of $\y_t$ given $\x_t$. When this function is used as a likelihood, we write $g_t^{\y_t}(\x_t):=g_t(\y_t,\x_t)$ to emphasize that it is a function of $\x_t$.
\item Let $f:\mZ \rw \Real$ be a real function on some set $\mZ$. We denote the absolute supremum of $f$ as $\| f \|_\infty := \sup_{\z \in \mZ} |f(\z)|$. The set of bounded real functions on $\mZ$ is $B(\mZ) := \{ f:\mZ \rw \Real \mbox{ such that } \|f\|_\infty < \infty \}$.  
\item Let $\bfa = (a_1, ..., a_d)$ be a multi-index, where each $a_i$, $i=1, 2, ..., d$, is a non-negative integer. Let $f : \mZ \rw \Real$ be a real function on a $d$-dimensional set $\mZ \subseteq \Real^d$. We use $D^\bfa f(\z)$ to denote the partial derivative of $f$ w.r.t. the variable $\z$ determined by the entries of $\bfa$, namely,
$$
D^\bfa f(\z) = \frac{
	\partial^{a_1} \cdots \partial^{a_d} f
}{
	\partial z_1^{a_1} \cdots \partial z_d^{a_d}
}(\z).
$$ 
The order of the derivative operator $D^\bfa$ is $|\bfa|=\sum_{i=1}^d a_i$.
\item The minimum out of two scalar quantities, $a,b \in \Real$, is denoted $a \wedge b$.
\end{itemize}

We make the following assumptions on the likelihood function $g_t$ and the predictive observation measure $\mu_t(d\y_t) = p_t(\y_t) d\y_t$.
\begin{itemize}
\item[\mfL] For each $t \ge 1$, the function $g_t$ is positive and bounded, i.e., $g_t(\y,\x)>0$ for any $(\y,\x) \in {\cal Y} \times {\cal X}$ and $\| g_t \|_\infty = \sup_{(\y,\x) \in \Y \times \X} | g_t(\y,\x) | <\infty$.

\item[\mfD] For each $t \ge 1$, the function $g_t(\y,\x)$ is differentiable with respect to $\y$, with bounded derivatives up to order $d_y$, hence $D^{\bf 1} g_t(\y,\x) = \frac{
	\partial^{d_y} g_t
}{
	\partial y_1 \cdots \partial y_{d_y}
}(\y,\x)$ exists and 
\begin{equation}
\| D^{\bf 1} g_t \|_\infty  = \sup_{(\y,\x) \in {\cal Y} \times {\cal X}} | D^{\bf 1} g_t(\y,\x) | <\infty.
\nonumber
\end{equation}

\item[\mfC] For any $0 < \beta < 1$ and any $p \ge 4$, the sequence of hypercubes 
\begin{equation}
C_M := \left[ -\frac{M^\frac{\beta}{p}}{2}, +\frac{M^\frac{\beta}{p}}{2} \right] \times \cdots \times \left[ -\frac{M^\frac{\beta}{p}}{2}, +\frac{M^\frac{\beta}{p}}{2} \right] \subset \Real^{d_y}
\nonumber
\end{equation}
satisfies the inequality $\mu_t(\overline{C_M}) \le b M^{-\eta}$ for some constants $b>0$ and $\eta>0$ independent of $M$ (yet possibly dependent on $\beta$ and $p$), where $\overline{C_M}=\Real^{d_y}\backslash C_M$ is the complement of $C_M$.
\end{itemize}

\begin{Nota}
Assumptions \mfL~and \mfD~refer to regularity conditions (differentiability and boundedness) that the likelihood function of the state space model should satisfy. Models of observations, for example, of the form $\y_t = f(\x_t) + \bfu_t$, where $f$ is a (possibly nonlinear) transformation of the state $\x_t$ and $\bfu_t$ is noise with some differentiable, exponential-type pdf (e.g., Gaussian or mixture-Gaussian), readily satisfy these assumptions. Typical two-sided heavy-tailed distributions, such as Student's $t$ distribution, also satisfy \mfL~and \mfD.

%Assumption \mfC~states that the tails of the pdf $p_t(\y_t)=p(\y_t|\y_{1:t-1})$ should not be too heavy. Being polynomial on $M$, this constraint is relatively weak and the assumption is satisfied for all exponential-type distributions as well as for many heavy-tailed distributions. For example, when $d_y=1$, one can choose the constants $b$ and $\eta$ such that $bM^{-\eta}$ is an upper bound for the tails of the (heavy-tailed) Pareto, Weibull, Burr or Levy distributions.
\end{Nota}

\begin{Nota}
Assumption \mfC~states an explicit bound on the probability under the tails of the pdf $p_t(\y_t)=p(\y_t|\y_{1:t-1})$. The bound is polynomial, namely
$$
\mu_t(\overline{C_M}) = 1 - \int_{-\frac{1}{2}M^{\frac{\beta}{p}}}^{\frac{1}{2}M^{\frac{\beta}{p}}} \cdots \int_{-\frac{1}{2}M^{\frac{\beta}{p}}}^{\frac{1}{2}M^{\frac{\beta}{p}}} p_t(\y) d\y \le b M^{-\eta},
$$
and therefore immediately verified, e.g., by all distributions of the exponential family as well as for many heavy-tailed distributions. For example, when $d_y=1$ (i.e., the observations are scalars), one can choose the constants $b$ and $\eta$ such that $bM^{-\eta}$ is an upper bound for the tails of the (heavy-tailed) Pareto, Weibull, Burr or Levy distributions.\\
It is actually possible to find simple conditions on the conditional pdf of the observations, $g_t(\y_t,\x_t)$, that turn out sufficient for assumption \mfC~to hold true. Let us keep $d_y=1$, for simplicity, and assume that there exists a sequence of positive constants $\{c_t\}_{t \ge 1}$ such that $g_t(y_t,\x_t)$ has a polynomial upper bound itself, namely
\begin{equation}
\sup_{\x_t \in \mX} g_t(y_t,\x_t) \le c_t |y_t|^{-(1+\epsilon)}
\label{eqRev0}
\end{equation}
for some $\epsilon>0$ and every $y_t$ such that $|y_t| > \frac{1}{2}$ (note that the smallest set in the sequence $C_M$ is $C_1=\left[-\frac{1}{2},\frac{1}{2}\right]$). For probability distributions with infinite support and continuous with respect to the Lebesgue measure, the inequality \eqref{eqRev0} implies that the densities $g_t(y_t,\x_t)$ are integrable for every possible choice of $\x_t \in \mX$. Then, the probability below the right tail of $p_t(y)$ is
\begin{eqnarray}
\int_{\frac{1}{2}M^{\frac{\beta}{p}}}^\infty p_t(y)dy &=& \int_{\frac{1}{2}M^{\frac{\beta}{p}}}^\infty \int_{\mX} g_t(y,\x) \tilde p_t(\x) d\x dy \nonumber\\
&\le& c_t \int_{\frac{1}{2}M^{\frac{\beta}{p}}}^\infty y^{-(1+\epsilon)} \int_{\mX} \tilde p_t(\x) d\x dy, \nonumber
\end{eqnarray} 
where the inequality follows from the application of \eqref{eqRev0}. Since $\tilde p_t(\x)$ is a pdf, we have $\int_{\mX} \tilde p_t(\x) d\x=1$ and some elementary calculations yield
\begin{eqnarray}
\int_{\frac{1}{2}M^{\frac{\beta}{p}}}^\infty p_t(y)dy &\le& c_t \int_{\frac{1}{2}M^{\frac{\beta}{p}}}^\infty y^{-(1+\epsilon)} dy \nonumber\\
&=& c_t \left[
	-\frac{y^{-\epsilon}}{\epsilon}
\right]_{\frac{1}{2}M^{\frac{\beta}{p}}}^\infty
  = \frac{2^\epsilon c_t}{\epsilon} M^{-\frac{\epsilon\beta}{p}}.
\label{eqRev1}
\end{eqnarray}
The same result is easily obtained for the left tail of $p_t(y)$, hence
\begin{eqnarray}
\mu_t(\overline{C_M}) &=& \int_{\frac{1}{2}M^{\frac{\beta}{p}}}^\infty p_t(y)dy + \int^{-\frac{1}{2}M^{\frac{\beta}{p}}}_{-\infty} p_t(y)dy \nonumber\\
&\le& \frac{2^{1+\epsilon} c_t}{\epsilon} M^{-\frac{\epsilon\beta}{p}}. \label{eqRev2}
\end{eqnarray}
By comparing \eqref{eqRev2} and the inequality $\mu_t(\overline{C_M}) \le b M^{-\eta}$, we readily see that we can choose $b=\frac{2^{1+\epsilon} c_t}{\epsilon}$ and $\eta=\frac{\epsilon\beta}{p}>0$ to uphold assumption \mfC. A similar derivation can be carried out when $d_y > 1$.
\end{Nota}

\begin{Teorema} \label{th1}
Assume that \mfL, \mfD~and \mfC~hold and the observations $\y_{1:t-1}$ are fixed (and otherwise arbitrary). Then, for every $h \in B({\cal Y})$ and any $\epsilon \in (0,\frac{1}{2})$ there exists an a.s. finite r.v. $W_t^\epsilon$, independent of $M$, such that
\begin{equation}
\left| ( h, \mu_t^M ) - ( h, \mu_t ) \right| \leq \frac{W_t^\epsilon}{M^{(\frac{1}{2}-\epsilon) \wedge \eta}}.
\nonumber
\end{equation}
In particular, 
$$
\lim_{M\rw\infty} ( h, \mu_t^M ) = (h,\mu_t) \quad \mbox{a.s.}
$$
\end{Teorema}

See Appendix \ref{apTh1} for a proof.

Note that the r.v. $W_t^\epsilon$ in the statement of Theorem \ref{th1} depends on the time instant $t$. It is possible to remove this dependence if the constants $b$ and $\eta$ in assumption \mfC~are chosen to be independent of $t$ and we impose further constraints on the likelihood function and the Markov kernel of the state space model (similar to the sufficient conditions for uniform convergence in, e.g., \cite{DelMoral04} or \cite{Heine08}).

%%%%%%%%%%%%%%%%%%%%%%
%%%%%%%%%%%%%%%%%%%%%%
%%%%%%%%%%%%%%%%%%%%%%
\section{Online selection of the number of particles}
\label{sec_method}
%%%%%%%%%%%%%%%%%%%%%%
%%%%%%%%%%%%%%%%%%%%%%
%%%%%%%%%%%%%%%%%%%%%%

In the sequel we assume scalar observations, hence $d_y=1$ and $\y_t=y_t$ (while $d_x \ge 1$ is arbitrary). A discussion of how to proceed when $d_y > 1$ is provided in Section \ref{ssMultidim}. 
 
Our goal is to evaluate the convergence of the BPF (namely, the accuracy of the approximation $p_t^M(y_t)$) in real time and, based on the convergence assessment, adapt the computational effort of the algorithm, i.e., the number of used particles $M$.

To that end, we run the BPF in the usual way with a light addition of computations. At each iteration we generate $K$ ``fictitious observations'', denoted $\tilde y_t^{(1)}, \ldots, \tilde y_t^{(K)}$, from the approximate predictive pdf $p_t^M(y_t)$. If the BPF is operating with a small enough level of error, then Theorem \ref{th1} states that these fictitious observations come approximately from the same distribution as the acquired observation, i.e., $\mu_t^M(dy_t) \approx \mu_t(dy_t)$. In that case, as we explain in Subsection \ref{sec_inv_sta},  a statistic $a_t^K$ can be constructed using $y_t, \tilde y_t^{(1)}, \ldots, \tilde y_t^{(K)}$, which necessarily has an (approximately) uniform distribution independently of the specific form of the state-space model \eqref{eqPrior}--\eqref{eqLikelihood}. By collecting a sequence of such statistics, say $a_{t-W+1}^K, \ldots, a_t^K$ for some window size $W$, one can easily test whether their empirical distribution is close to uniform using standard procedures. The better the approximation $\mu_t^M \approx \mu_t$ generated by the BPF, the better fit with the uniform distribution can be expected.

If $K <<M$ and $W$ is not too large, the cost of the added computations is negligible compared to the cost of running the BPF with $M$ particles and, as we numerically show in Section \ref{section_numerical}, the ability to adapt the number of particles online leads to a very significant reduction of the running times without compromising the estimation accuracy. 

Below we describe the method, justify its theoretical validity and discuss its computational complexity as well as its extension to the case of multidimensional $\y_t$'s.

%%%%%%%%%%%%%%%%%%%%%%
\subsection{Generation of fictitious observations}
\label{sec_fict_obs}
%%%%%%%%%%%%%%%%%%%%%%

The proposed method demands at each time $t$ the generation of $K$ fictitious observations (i.e., Monte Carlo samples), denoted $\{ \tilde y_t^{(k)} \}_{k=1}^K$, from the approximate predictive observation pdf $p_t^M(y_t) = \frac{1}{M}\sum_{m=1}^M p(y_t|\bar \x_t^{(m)})$. Since the latter density is a finite mixture, drawing from $p_t^M(y_t)$ is straightforward as long as the conditional density of the observations, $p(y_t|\x_t)$, is itself amenable to sampling. In order to generate $\tilde y_t^{(k)}$, it is enough to draw a sample $j^{(k)}$ from the discrete uniform distribution on $\{ 1, 2, ..., M \}$ and then generate $\tilde y_t^{(k)} \sim p(y_t|\bar \x_t^{(j^{(k)})})$.

%%%%%%%%%%%%%%%%%%%%%%
\subsection{Assessing convergence via invariant statistics}
%%%%%%%%%%%%%%%%%%%%%%
\label{sec_inv_sta}
For simplicity, let us assume first that $p_t^M(y_t) = p_t(y_t) = p(y_t|y_{1:t-1})$, i.e., there is no approximation error and, therefore, the fictitious observations $\{ \tilde y_t^{(k)} \}_{k=1}^K$ have the same distribution as the {\em true} observation $y_t$. We define the set $\mA_{K,t} := \{Â y \in \{ \tilde{y}_t^{(k)} \}_{k=1}^K : y < y_t \}$ and the r.v. $A_{K,t} := | \mA_{K,t} | \in \{0,1, ..., K \}$. Note that $\mA_{K,t}$ is the set of fictitious observations which are smaller than the actual one, while $A_{K,t}$ is the number of such observations. If we let $\mbbQ_K$ denote the probability mass function (pmf) of $A_K$, it is not hard to show that $\mbbQ_K$ is uniform independently of the value and distribution of $y_t$. This is rigorously given by the Proposition below.

%%%
\begin{Proposicion} \label{prop_pmf}
If $y_t, \tilde{y}_t^{(1)}, \ldots, \tilde y_t^{(K)}$ are i.i.d. samples from a common continuous (but otherwise arbitrary) probability distribution, then the pmf of the r.v. $A_{K,t}$ is
\beq
\label{pmf_uniform}
\mbbQ_{K}(n) = \frac{1}{K+1}, \qquad n=0,...,K.
\eeq
\end{Proposicion}

\noindent \textit{\textbf{Proof}}: Since $y_t, \tilde{y}_t^{(1)}, \cdots,\tilde{y}_t^{(K)}$ are i.i.d., all possible orderings of the $K+1$ samples are a priori equally probable, and the value of the r.v. $A_{K,t}$ depends uniquely on the relative position of $y_t$ after the samples are sorted  (e.g., if $y_t$ is the smallest sample, then $A_{K,t}=0$, if there is exactly one $\tilde{y}_t^{(i)} < y_t$ then $A_{K,t}=1$, etc.). There are $(K+1)!$ different ways in which the samples $y_t, \tilde{y}_t^{(1)}, \cdots, \tilde{y}_t^{(K)}$ can be ordered, but $A_{K,t}$ can only take values from $0$ to $K$. In particular, given the relative position of $y_t$, there are $K!$ different ways in which the remaining samples $\tilde{y}_t^{(1)}, \cdots, \tilde{y}_t^{(K)}$ can be arranged. Therefore, $\mbbQ_K(A_K = n)=\frac{K!}{(K+1)!}=\frac{1}{K+1}$ for every $n\in\{0,1,...,K\}$. \qed

For the case of interest in this paper, the r.v.'s $y_t, \tilde{y}_t^{(1)}, \ldots, \tilde y_t^{(K)}$ (the actual and fictitious observations) have a common probability distribution given by the measure $\mu_t$ and are generated independently. For the class of state space models described in Section \ref{sPF}, and the explicit assumptions in Section \ref{sec:convergence}, the measure $\mu_t$ is absolutely continuous w.r.t. the Lebesgue measure (with associated density $p_t(y)$) and, therefore, $y_t, \tilde{y}_t^{(1)}, \ldots, \tilde y_t^{(K)}$ are indeed continuous r.v.'s and the assumptions of Proposition \ref{prop_pmf} are met. Moreover, it can also be proved that the variables in the sequence $A_{K,t}$ are independent.
%%%
\begin{Proposicion} \label{prop_indep}
If the r.v.'s $y_t, \tilde{y}_t^{(1)}, \ldots, \tilde y_t^{(K)}$ are i.i.d. with common pdf $p_t(y)$, then the r.v.'s in the sequence $\{ A_{K,t} \}_{t \ge 1}$ are independent.
\end{Proposicion}
See Appendix \ref{appendix_prop_pmf} for a proof.

In practice, $p_t^M(y_t)$ is just an approximation of the predictive observation pdf $p_t(y_t)$ and, therefore, the actual and fictitious observations are not i.i.d. However, under the assumptions of Theorem \ref{th1}, the a.s. convergence of the approximate measure $\mu_t^M(dy_t)=p_t^M(y_t)dy_t$ enables us to obtain an ``approximate version'' of the uniform distribution in Proposition \ref{prop_pmf}, with the error vanishing as $M\rw\infty$. To be specific, we introduce the set $\mA_{K,M,t} := \{Â y \in \{ \tilde{y}_t^{(k)} \}_{k=1}^K : y < y_t \}$, which depends on $M$ because of the mismatch between $p_t^M(y_t)$ and $p_t(y_t)$, and the associated r.v. $A_{K,M,t}=|\mA_{K,M,t}|$ with pmf $\mbbQ_{K,M,t}$. We have the following convergence result for $\mbbQ_{K,M,t}$.

%%%%%%%%%%%%%%

\begin{Teorema} \label{thQM}
Let $y_t$ be a sample from $p_t(y_t)$ and let $\{ \tilde{y}_t^{(k)} \}_{k=1}^K$ be i.i.d. samples from $p_t^M(y_t)$. If the observations $y_{1:t-1}$ are fixed and Assumptions \mfL, \mfD~and \mfC~hold, then there exists a sequence of non-negative r.v.'s $\{\varepsilon_t^M\}_{M \in \mathbb{N}}$ such that $\lim_{M\rw\infty} \varepsilon_t^M = 0$ a.s. and
\begin{equation} 
\frac{1}{K+1} - \varepsilon_t^M \leq \mbbQ_{K,M,t}(n) \leq \frac{1}{K+1} + \varepsilon_t^M.
\label{eqGG}
\end{equation}
In particular, $\lim_{M\rw\infty} \mbbQ_{K,M,t}(n) = \mbbQ_K(n) = \frac{1}{K+1}$ a.s.
\end{Teorema}

See Appendix \ref{apThQM} for a proof. Proposition \ref{prop_pmf} states that the statistic $A_{K,t}$ is distribution-invariant, since $\mbbQ_K(n)=\frac{1}{K+1}$ independently of $t$ and the state space model. Similarly, Theorem \ref{thQM} implies that the statistic $A_{K,M,t}$ is asymptotically distribution-invariant (independently of $t$ and the model) since $\mbbQ_{K,M,t}(n) \rw \frac{1}{K+1}$ when $M\rw\infty$, as the BPF converges.\footnote{Specifically note that, under assumptions \mfL, \mfD~and \mfC, the convergence of the continuous random measure $\mu_t^M$ computed via the BPF (which is sufficient to obtain \eqref{eqGG}; see Appendix \ref{apThQM}) is guaranteed by Theorem \ref{th1}.}

%%%%%%%%%%%%%%%%%%%%%%
\subsection{Algorithm with adaptive number of particles }
\label{sec_algorithm}
%%%%%%%%%%%%%%%%%%%%%%

We propose an algorithm that dynamically adjusts the number of particles of the filter based on the transformed r.v. $A_{K,M,t}$. 
Table \ref{table_algorithm} summarizes the proposed algorithm, that is  embedded into a standard BPF (see Section \ref{sec_back}) but can be applied to virtually any other particle filter in a straightforward manner.
The parameters of the algorithm are shown in \mbox{Table \ref{table_parameters}}.

The BPF is initialized in Step 1(a) with $M_0$ initial particles.
At each recursion, in Step 2(a), the filtered distribution of the current state is approximated.
In Step 2(b), $K$ fictitious observations $\{\tilde{y}_t^{(k)}\}_{k=1}^K$ are drawn and the statistic $A_{K,M,t} = a_{K,M,t}$ is computed.
In Step 2(c), once a set of $W$ consecutive statistics have been acquired, $\mathcal{S}_t=\{a_{K,M,t-W+1}, a_{K,M,t-W+2},...,a_{K,M,t-1},a_{K,M,t}\}$, a statistical test is performed for checking whether  $\mathcal{S}_t$ is a sequence of samples from the uniform pmf given by Eq. \eqref{pmf_uniform}.

There are several approaches that can be used to exploit the information contained in $\mathcal{S}_t$. Here we perform a Pearson's chi-squared test \cite{plackett1983karl}, where
the $\chi_t^2$ statistic is computed according to Eq. \eqref{chi_statistic} (see Table \ref{table_algorithm}).
Then, a p-value $p_{K,t}^*$ for testing the hypothesis that the empirical distribution of ${\cal S}_t$ is uniform is computed. 
The value $p_{K,t}^*$ is obtained by comparing the $\chi_t^2$ statistic with the $\chi^2$ distribution with $K$ degrees of freedom. Intuitively, a large $p_{K,t}^*$ suggests a good match of the sequence ${\cal S}_t$ with an i.i.d. sample from the uniform distribution on $\{0,1,..., K\}$, while a small $p_{K,t}^*$ indicates a mismatch.
Therefore, the p-value $p_{K,t}^*$ is compared with two different significance levels: a low threshold $p_\ell$ and a high threshold $p_h$.
If $p_{K,t}^* \leq p_\ell$, the number of particles is increased according to the rule $M_t = f_{\text{up}}(M_{t-1})$ whereas, if $p_{K,t}^*\geq p_h$, the number of particles is decreased according to the rule $M_t = f_{\text{down}}(M_{t-1})$.
When  $p_\ell  <  p_{K,t}^*  < p_h$, the number of particles remains fixed.
These two significance levels allow the practitioner to select the operation range by considering a performance-to-computational-cost tradeoff.
Note that we set $M_{\text{min}}$ and $M_{\text{max}}$, maximum and minimum values for the number of particles, respectively.

A large window $W$ yields a more accurate convergence assessment but increases the latency (or decreases the responsiveness) of the algorithm.
If the algorithm must be run online, this latency can be critical for detecting a malfunction of the filter and adapting consequently the number of particles. Therefore there is a tradeoff between the accuracy of the convergence assessment procedure and latency of the algorithm.

\begin{table}[!t]
	\centering
	\caption{\textbf{Parameters of the algorithm}}
	\begin{tabular}{|p{0.95\columnwidth}|}
    \hline
		\footnotesize
\begin{itemize}[label={--}]
\item $M_0$, initial number of particles
\item $M_{\text{min}}$, minimum number of particles
\item $M_{\text{max}}$, maximum number of particles
\item $K$, number of fictitious samples per iteration
\item $W$, window length
\item $p_{\ell}$, lower significance level of p-values
\item $p_{h}$, higher significance level of p-values
\item $f_{\text{up}}(\cdot)$, rule for increasing $M$
\item $f_{\text{down}}(\cdot)$, rule for decreasing $M$
\end{itemize}\\
		\hline
\end{tabular}
\label{table_parameters}
\end{table}

\begin{table}[!t]
	\centering
	\caption{\textbf{Algorithm for adapting the number of particles}}
	\begin{tabular}{|p{0.95\columnwidth}|}
    \hline
		\footnotesize

\begin{enumerate}
\item {\bf [Initialization]}
\begin{enumerate}
\item Initialize the particles and the weights of the filter as $$\x_0^{(m)}\sim p(\x_0),\quad\quad m=1,\ldots,M_0,$$  $$w_0^{(m)}=1/M_0,\quad\quad m=1,\ldots,M_0,$$
\end{enumerate}
and set $n=1$.
\item {\bf [For $t=1:T$]}
\begin{enumerate}

\item {\bf Bootstrap particle filter:}
\begin{itemize}[label={--}]
\item Resample $M_n$ samples of $\bar{\x}_{t-1}^{(m)} $ with weights $w_{t-1}^{(m)}$ to obtain $\x_{t-1}^{(m)}$.
\item Propagate $\bar{\x}_t^{(m)}\sim p(\x_t|\x_{t-1}^{(m)}),\quad\quad m=1,\ldots,M_n$.
\item Compute the non-normalized weights $\bar{w}_{t}^{(m)}=p(y_t|\bar{\x}_t^{(m)}),$ $\quad\quad m=1,\ldots,M_n$.
\item Normalize the weights $\bar{w}_{t}^{(m)}$ to obtain ${w}_{t}^{(m)}$, $\quad\quad m=1,\ldots,M_n$. 

\end{itemize}
\item {\bf Fictitious observations:}
\begin{itemize}[label={--}]
\item Draw $\tilde{y}_t^{(k)} \sim p^M(y_t|y_{t-1}), k=1,\ldots,K$. 
\item Compute $a_{K,M,t}=A_{K,M,t}$, i.e., the position of $y_t$ within the set of ordered fictitious observations $\{ \tilde{y}_t^{(k)} \}_{k=1}^{K}$.
\end{itemize}
\item If $t=nW$, 
(\textbf{assessment of convergence}):
\begin{itemize}[label={--}]
\item Compute the $\chi_t^2$ statistic over the empirical distribution of $\mathcal{S}_t=\{a_{K,M,t}, a_{K,M,t-1},...,a_{K,M,t-W+1}\}$ as

\begin{equation}
\label{chi_statistic}
\chi_t^2 = \sum_{j=0}^K \frac{(O_j - E_j)^2}{E_j},
\end{equation}
where $O_j$ is the frequency of the observations in the window being in the $j$th relative position, i.e., $O_j=|a_{K,M,\tau} \in \mathcal{S}_t: a_{K,M,\tau} = j|$, and $E_j$ is the expected frequency under the null hypothesis, i.e., $E_j=W\cdot \mbbQ_{K}(j) =\frac{W}{K+1}$ (see Eq. \eqref{pmf_uniform}).

\item Calculate the p-value $p_{K,t}^*$  by comparing the statistic $\chi_t^2$  to the $\chi^2$-distribution with $K$ degrees of freedom.
\item If $p_{K,t}^* \leq p_{\ell}$
\begin{itemize}[label={}]
\item increase $M_n = \min\{f_{\text{up}}(M_{n-1}),M_{\text{max}}\}$.
\end{itemize}
\item Else, if $p_{K,t}^* \geq p_{h}$,
\begin{itemize}[label={}]
\item decrease $M_n = \max\{f_{\text{down}}(M_{n-1}),M_{\text{min}}\}$.
\end{itemize}
\item Else,
\begin{itemize}[label={}]
\item $M_n = M_{n-1}$.
\end{itemize}
\item Set $n=n+1$.
\end{itemize}
\item If $t<Wn$, set $t=t+1$ and go to $2$. Otherwise, end.
\end{enumerate}

\end{enumerate} \\
		\hline
\end{tabular}
\label{table_algorithm}
\end{table}

%%%%%%%%%%%%%%%%%%%%%%
\subsection{Computational cost}
%%%%%%%%%%%%%%%%%%%%%%

Compared to the BPF, the additional computational cost of the method is mainly driven by the generation of the $K$ fictitious observations at each iteration as shown in Subsection \ref{sec_fict_obs}.
The generation of these fictitious observations is a two-step procedure, where in the first step, we draw $K$ discrete indices, say $j_1, ..., j_K$, from the set $\{ 1, ..., M_n \}$ with uniform probabilities, and in the second step, we draw  $K$ samples from $p(y_t|\bar \x_t^{(j_1)}), \ldots, p(y_t|\bar \x_t^{(j_K)})$, respectively. 

In the proposed algorithm, a Pearson's $\chi^2$ test is performed with a sequence $\mathcal{S}_t$ of $W$ samples, that is, it is carried out only once every $W$ consecutive time steps.
Therefore, the computational cost will depend on the parameters $K$ and $W$. We will show in Section \ref{section_numerical} that the algorithm can work very well with a low number of fictitious observations, which imposes a very light extra computational load.

%%%%%%%%%%%%%%%%%%%%%%
\subsection{Multidimensional observations} \label{ssMultidim}
%%%%%%%%%%%%%%%%%%%%%%
Through this section, we have assumed scalar observations. In the multidimensional case, with $\y_t=[y_{1,t}, \ldots, y_{d_y,t}]^\top$, the same assessment scheme can be applied over each marginal $p(y_{i,t}|\y_{1:t-1})$ of the predictive observation pdf. Theoretical guarantees readily follow from the convergence of the marginal measures $\mu_{i,t}^M(dy_{i,t})=p^M(y_{i,t}|\y_{1:t-1})dy_{i,t}$ under the same assumptions as the joint measure $\mu_t^M$ (see Appendix \ref{apTh1}).

The algorithm proposed in Section \ref{sec_algorithm} can be extended to the case with multidimensional observations. One of way of doing it is by performing an independent assessment for each marginal pdf $p(y_{i,t}|\y_{1:t-1})$. As a result, $d_y$ p-values $p_{K,t,i}^{*}$, with $i=1,...,d_y$, become available for deciding whether to increase, decrease or keep fixed the number of particles. A conservative approach is to increase the number of particles whenever at least one p-value $p_{K,t,i}^{*}$ is below the threshold $p_\ell$. Note that the complexity of this approach grows with the dimension of the observations.

Finally, note that the convergence of the marginals does not imply the convergence of the joint approximation $\mu_t^M$. However, it can be reasonably expected that when all the marginals are approximated well over a period of time, the joint distribution is accurately approximated as well.

%%%%%%%%%%%%%%%%%%%%%%
%%%%%%%%%%%%%%%%%%%%%%
%%%%%%%%%%%%%%%%%%%%%%
\section{Numerical example}
\label{section_numerical}
%%%%%%%%%%%%%%%%%%%%%%
%%%%%%%%%%%%%%%%%%%%%%
%%%%%%%%%%%%%%%%%%%%%%

%%%%%%%%%%%%%%%%%%%%%%
\subsection{The three-dimensional Lorenz system} \label{ssLorenz}
\label{model_lorenz}
%%%%%%%%%%%%%%%%%%%%%%
\subsubsection{Model description}
In this section we show computer simulation results that demonstrate the performance of the proposed method.
 We consider the problem of tracking the state of a three-dimensional Lorenz system \cite{Lorenz63} with additive dynamical noise, partial observations and additive measurement noise \cite{Chorin04}.
 Namely, we consider a three-dimensional stochastic process $\{ \X(s) \}_{s\in(0,\infty)}$ taking values on $\Real^3$, whose dynamics are described by the system of stochastic differential equations
\begin{eqnarray}
dX_1 &=& -{\sf s} (X_1-Y_1) + dW_1, \nonumber\\
dX_2 &=& {\sf r} X_1 - X_2 - X_1X_3 + dW_2, \nonumber\\
dX_3 &=& X_1X_2 - {\sf b} X_3 + dW_3, \nonumber
\end{eqnarray}
where $\{ W_i(s) \}_{s\in(0,\infty)}$, $i=1, 2, 3$, are independent one-dimensional Wiener processes and
$$
({\sf s,r,b}) = \left(
	10, 28, \frac{8}{3}
\right)
$$
are static model parameters broadly used in the literature since they lead to a chaotic behavior \cite{Lorenz63}. Here we use a discrete-time version of the latter system using an Euler-Maruyama scheme with integration step $\Delta = 10^{-3}$, which yields the model
\begin{eqnarray}
X_{1,n} &=& X_{1,n-1} - \Delta {\sf s} (X_{1,n-1}-X_{2,n-1}) + \sqrt{\Delta} U_{1,n},\label{eqDiscreteLorenz-1}\\
X_{2,n} &=& X_{2,n-1} + \Delta ( {\sf r} X_{1,n-1} - X_{2,n-1} - X_{1,n-1}X_{3,n-1} ) \nonumber \\ && + \sqrt{\Delta} U_{2,n}, \label{eqDiscreteLorenz-2}\\
X_{3,n} &=& X_{3,n-1} + \Delta ( X_{1,n-1}X_{2,n-1} - {\sf b} X_{3,n-1} ) \nonumber \\ && + \sqrt{\Delta} U_{3,n}, \label{eqDiscreteLorenz-3}
\end{eqnarray}
where $\{ U_{i,n} \}_{n=0, 1, ...}$, $i=1,2,3$, are independent sequences of i.i.d. normal random variables with zero mean and unit variance. The system \eqref{eqDiscreteLorenz-1}-\eqref{eqDiscreteLorenz-3} is partially observed every 200 discrete-time steps. Specifically, we collect a sequence of scalar observations $\{ Y_t \}_{t=1, 2, ...}$, of the form
\begin{equation}
Y_t = X_{1,200t} + V_t,
\label{eqObservLorenz}
\end{equation}
where the observation noise $\{ V_t \}_{t=1, 2, ...}$ is a sequence of i.i.d. normal random variables with zero mean and variance $\sigma^2 = \frac{1}{2}$.

Let $\X_n=(X_{1,n},X_{2,n},X_{3,n}) \in \Real^3$ be the state vector. The dynamic model given by Eqs. \eqref{eqDiscreteLorenz-1}--\eqref{eqDiscreteLorenz-3} defines the transition kernel $p(\x_{n}|\x_{n-1})$ and the observation model of Eq. \eqref{eqObservLorenz} is the likelihood function
$$
p(y_t|x_{1,200t})  \propto \exp\left\{ -\frac{1}{2\sigma^2}\left( y_t - x_{1,200t} \right)^2 \right\}.
$$
The goal is on tracking the sequence of joint posterior probability measures ${\pi_t}$, $t=1, 2, ...$, for $\{ \hat \X_t \}_{t=1, ...}$, where $\hat \X_t = \X_{200t}$. 
Note that one can draw a sample $\hat \X_t = \hat \x_t$ conditional on $\hat \X_{t-1} = \hat \x_{t-1}$ by successively simulating
$$
\tilde \x_n \sim p(\x_n|\tilde \x_{n-1}), \quad n=200(t-1)+1, ..., 200t,
$$
where $\tilde \x_{200(t-1)} = \hat \x_{t-1}$ and $\hat \x_t = \tilde \x_{200t}$. The prior measure for the state variables is normal, namely
$$
\X_0 \sim \mN(\x_*,v_0^2 \mI_3),
$$
where $\x_* = (-5.9165; -5.5233; 24.5723)$ is the mean and $v_0^2\mI_3$ is the covariance matrix of $\X_0$ , with $v_0^2 = 10$ and $\mI_3$ being the three-dimensional identity matrix.

%%%%%%%%%%%%%%%%%%%%%%
\subsubsection{Simulation setup} \label{lorenz_algorithm}
%%%%%%%%%%%%%%%%%%%%%%

With this example, we aim at showing how the proposed algorithm allows to operate the particle filter with a prescribed performance-to-computational-budget tradeoff.
With this purpose, we applied a standard BPF for tracking the sequence of posterior probability measures of the system system \eqref{eqDiscreteLorenz-1}-\eqref{eqDiscreteLorenz-3}   generated by the three-dimensional Lorenz model described above. %in Section \ref{ssLorenz}.
We generated a sequence of $T=2000$ synthetic observations, $\{ y_t; t=1, ..., 2000 \}$, spread over an interval of $400$ seconds (in continuous time), corresponding to $4 \times 10^5$ discrete time steps in the Euler-Maruyama scheme (hence, one observation every 200 steps).
Since the time scale of the discrete time approximation of Eqs. \eqref{eqDiscreteLorenz-1}--\eqref{eqDiscreteLorenz-3} is $n=200t$, a resampling step is taken every 200 steps of the underlying discrete-time system.

We started running the PF with a sufficiently large number of particles, namely $N=5000$, and then let the proposed algorithm decrease the number of particles to attain a prescribed point in the performance-to-computation-cost range.
This point is controlled by the operation range of the p-value, which is in turn driven by the pair of significance levels $[p_{\ell}-p_h]$.
We tested the algorithm for different ranges of p-values, namely, $p_{\ell} \in \{0.5, 0.4, 0.3, 0.2, 0.1, 0.05\}$ and $p_{h} \in \{0.9, 0.8, 0.7, 0.6, 0.5, 0.4, 0.3, 0.2, 0.1\}$.
When the p-value is below $p_{\ell}$, the algorithm doubles the number of particles $M_{n+1}= f_{\text{up}}(M_{n}) = 2M_{n}$, and when the p-value is over $p_h$, the number of particles is halved, $M_{n+1}=f_{\text{down}}(M_{n}) =M_{n}/2$.
We used $K=7$ fictitious observations and a window of size $W=20$.

In order to assess the approximation errors, we computed the empirical MSEs of the approximation of the posterior mean, $E[\hat \X_t | Y_{1:t}=y_{1:t}] $, by averaging the MSEs for the whole sequences. Note that, since the actual expectation cannot be computed in closed form for this system, we used the true underlying sequence $\{  \X_{200t} \}_{t=1, 2, ...}$ as the ground truth.

%%%%%%%%%%%%%%%%%%%%%%
\subsubsection{Numerical results} \label{ssResults}
%%%%%%%%%%%%%%%%%%%%%%

Table \ref{table_decreasing_2} shows results of the MSE of the approximation of the posterior mean, the average number of particles 
\begin{equation}
\label{M_mean}
\bar{M} = \frac{2}{T} \sum_{k=\frac{T}{2}+1}^{T} M_k,
\end{equation}
the p-values of the $\chi^2$ test, and the Hellinger distance \cite{nikulin2001hellinger} between the empirical distribution of $\mathcal{S}_t$ and the uniform distribution. They were obtained by averaging over $100$ runs and averaging over time for each run. The initial number of particles $M_0=2^{15}$, and the minimum and maximum number of particles are $M_{\text{min}}=2^{5}$ and $M_{\text{max}}=2^{15}$, respectively. The first half of time steps were discarded for obtaining the displayed results in order to test the behavior of the algorithm for different sets of parameters (see Eq. \eqref{M_mean}). Regarding the relation between the MSE and $\bar{M}$ and the p-values, it can be seen that selecting a high operation range yields good performance (low MSE) at the cost of using a large number of particles (high $\bar{M}$). When we decrease the range of p-values, the algorithm decreases the number of particles, increasing also the approximation error. Table \ref{table_decreasing_2} shows that this conclusion holds for any pair of $[p_{\ell}-p_h]$. 

\begin{figure}
\centering
\includegraphics[width=0.5\textwidth]{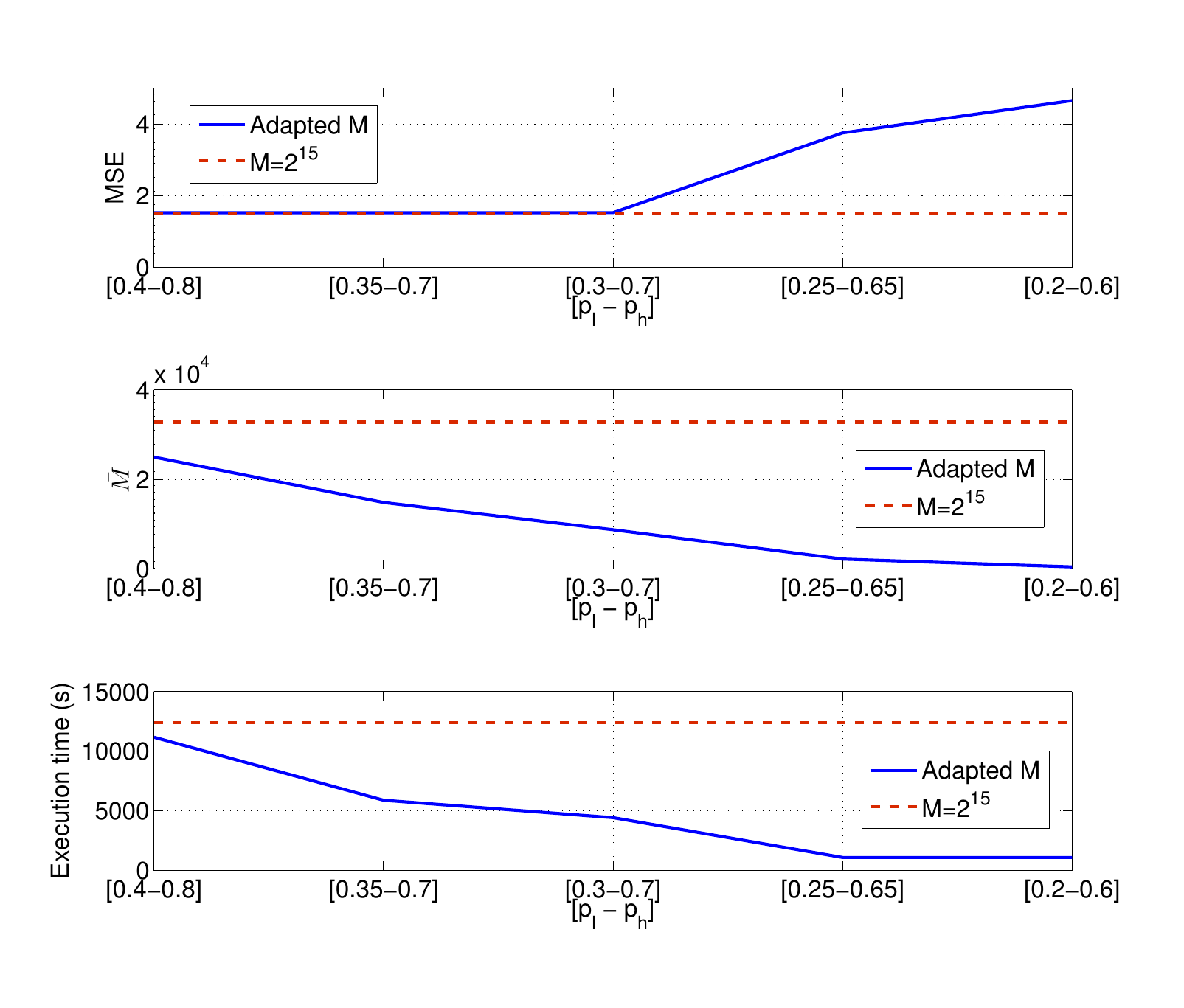}
\caption{Lorenz Model (Section \ref{model_lorenz}). MSE, number of particles $M$ and execution time for different pairs  of significance levels $[p_\ell - p_h]$ in solid blue line, and with a fixed number of particles $M=2^{15}$ in dashed red line.}
\label{fig_execution_time}
\end{figure}

Figure \ref{fig_execution_time} shows the MSE, the number of particles $\bar M$, and the execution time for the different operation ranges (solid blue line) compared to the particle filter with a fixed number of particles $M=2^{15}$ (dashed red line). It can be seen that with a moderate operation range ($[p_\ell - p_h]= [0.3 - 0.7]$), the algorithm can perform (in terms of MSE) similarly to the case with fixed $M$, while reducing the execution time approximately by a factor of four. The execution time can be further reduced by decreasing the operation range, although this worsens the performance.

\begin{figure}
\centering
\includegraphics[width=0.45\textwidth]{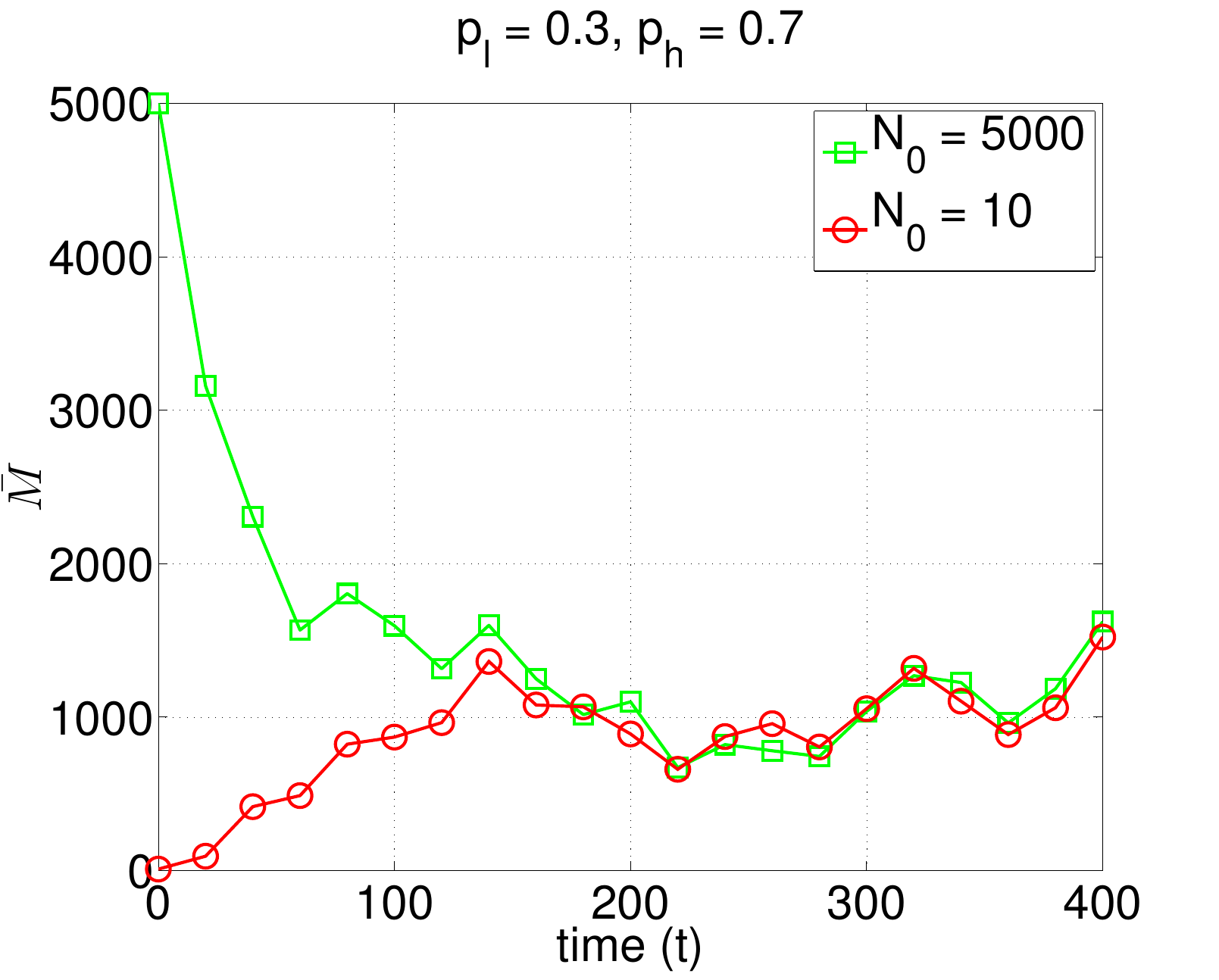}
\caption{Lorenz Model (Section \ref{model_lorenz}). Evolution of the number of particles adapted by the proposed algorithm when the initial number of particles $M_0 \in \{10, 5000 \}$. The significance levels were set to $p_\ell = 0.3$ and $p_h=0.7$.}
\label{fig_lorenz1}
\end{figure}

Figure \ref{fig_lorenz1} displays the evolution of the number of particles over time (averaged over $100$ runs) for $[p_\ell - p_h]= [0.3 - 0.7]$ both when $M_0=5000$ and $M_0 = 10$. In this case, the minimum and maximum number of particles are $M_{\text{min}}=10$ and $M_{\text{max}}=5000$, respectively. We see that, after some time, the number of particles adjusted by the algorithm does not depend on $M_0$. 

\begin{figure}
\centering
\includegraphics[width=0.45\textwidth]{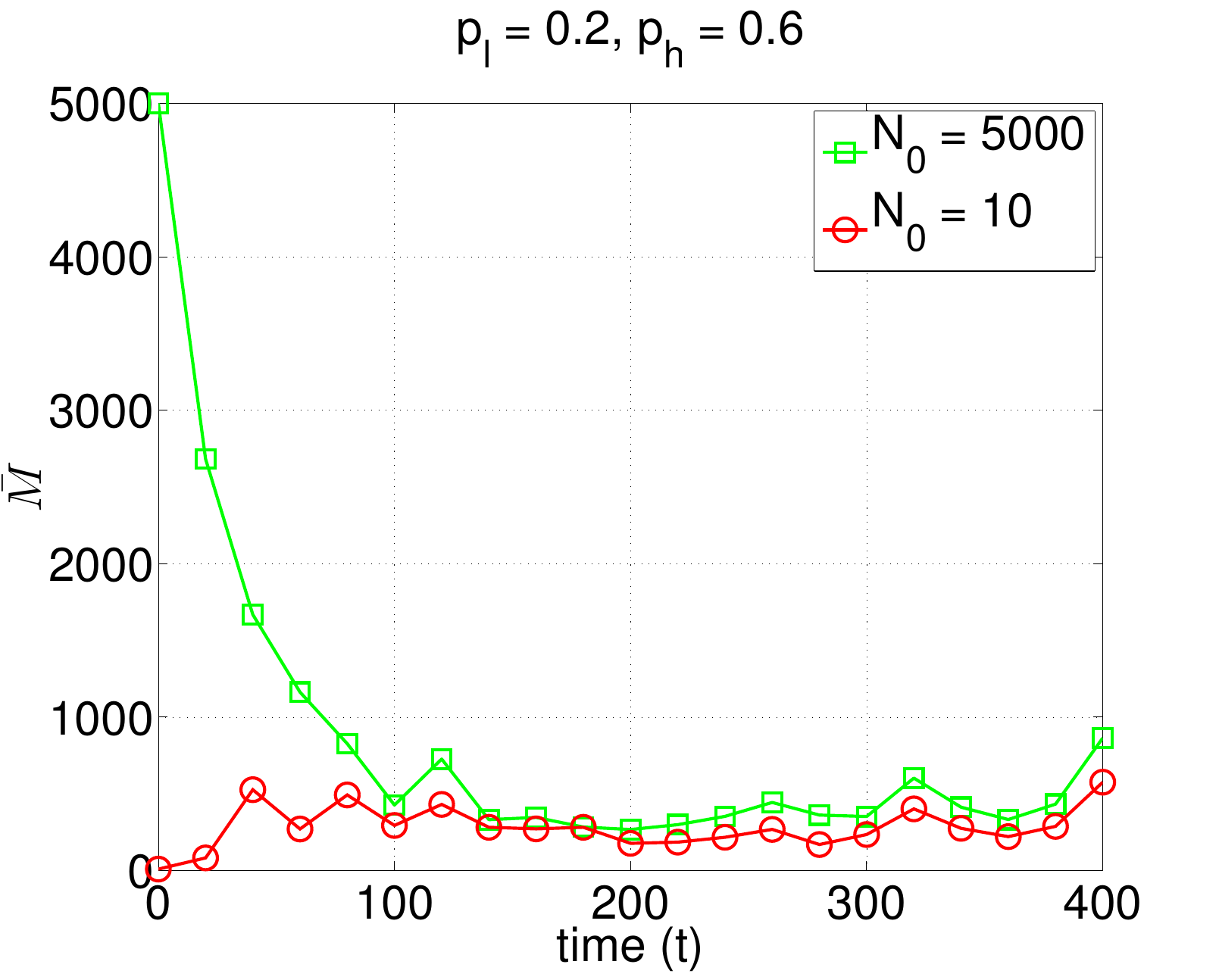}
\caption{Lorenz Model (Section \ref{model_lorenz}). Evolution of the number of particles adapted by the proposed algorithm when the initial number of particles $M_0 \in \{10,5000 \}$.  The significance levels were set to $p_\ell = 0.2$ and $p_h=0.6$.}
\label{fig_lorenz2}
\end{figure}

Figure \ref{fig_lorenz2}  shows the same behavior for  $[p_\ell - p_h]= [0.2 - 0.6]$. After some time, the filter uses less particles than the filter with results in Fig. \ref{fig_lorenz1} because the selected range of thresholds employs smaller p-values.

\begin{figure*}
\centering
\includegraphics[width=0.65\textwidth]{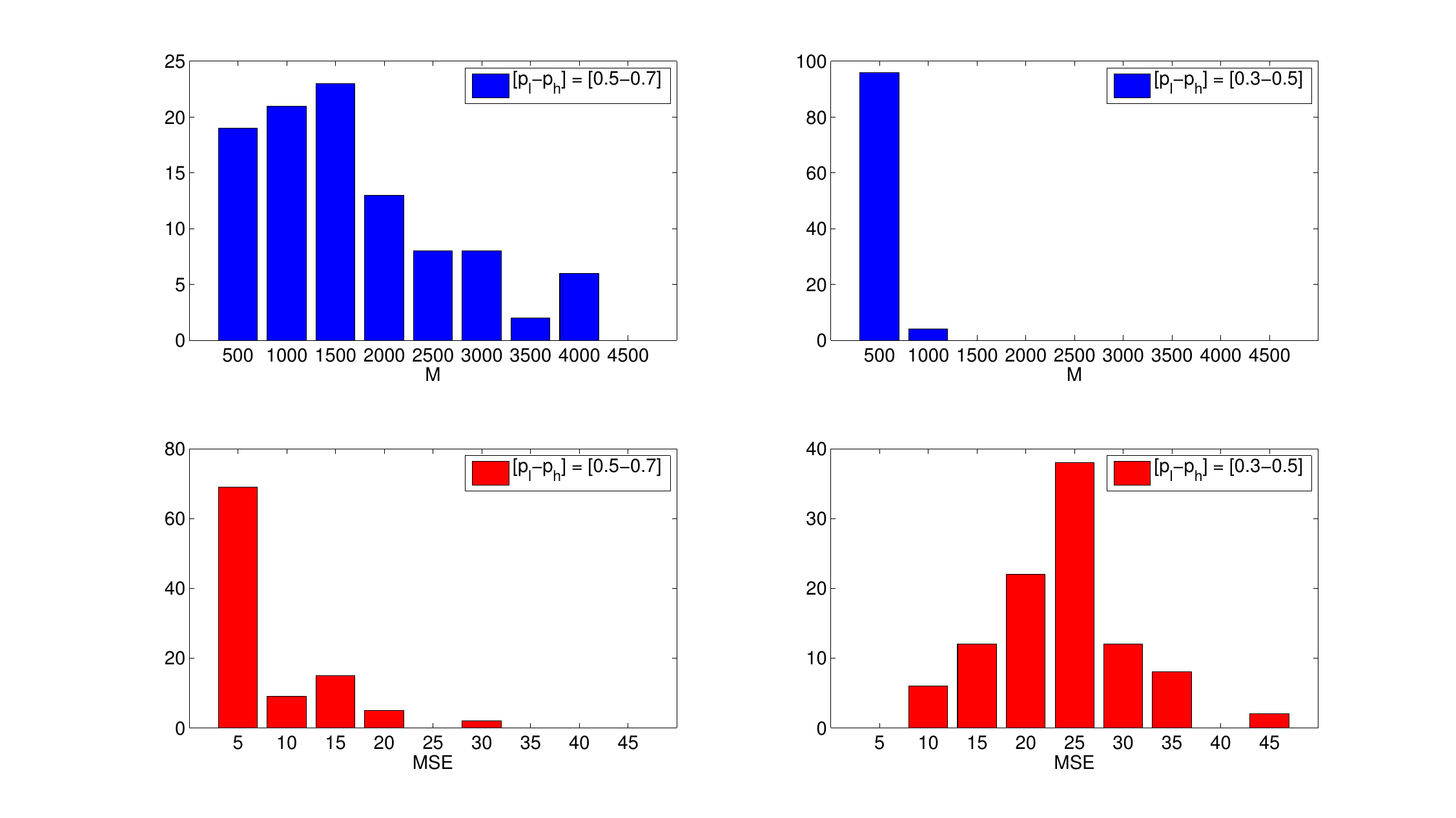}
\caption{Lorenz Model (Section \ref{model_lorenz}). Histograms of averaged MSE and M with $[p_\ell - p_h]= [0.3 - 0.5]$ and $[p_\ell - p_h]= [0.5 - 0.7]$. In both cases, the initial number of particles $M_0= 5000$.}
\label{fig_histogram}
\end{figure*}

\begin{figure}
\centering
\includegraphics[width=0.45\textwidth]{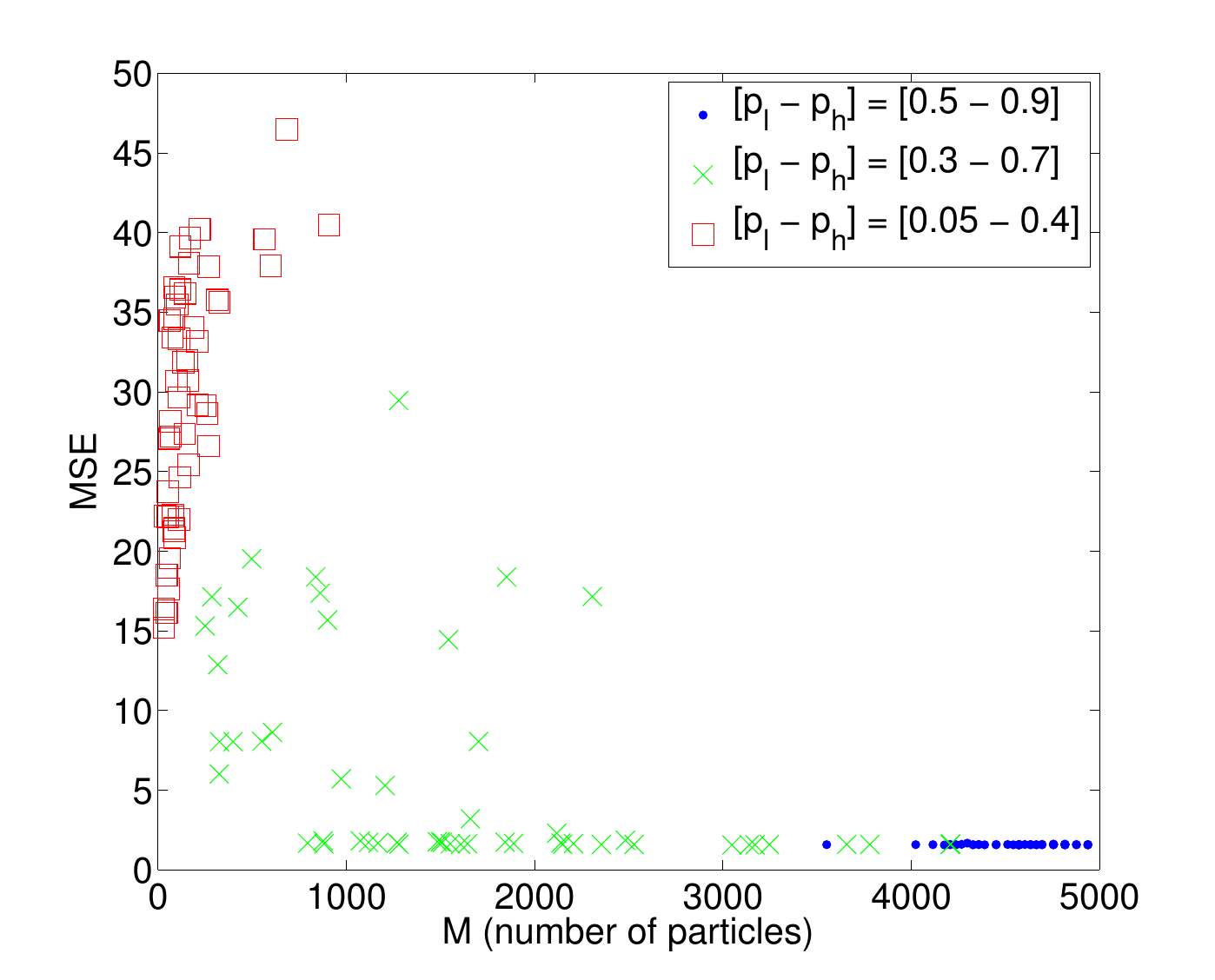}
\caption{Lorenz Model (Section \ref{model_lorenz}). MSE w.r.t. the averaged number of particles $M$ for runs with different sets of thresholds.}
\label{fig_MSE_M}
\end{figure}

Figure \ref{fig_histogram} shows histograms of averaged MSE and M for simulations performed with two different sets of thresholds: $[p_\ell - p_h]= [0.3 - 0.5]$ and $[p_\ell - p_h]= [0.5 - 0.7]$. In both cases, the initial number of particles is $M_0= 5000$. It can be seen that a more demanding pair of thresholds ($[p_\ell - p_h]= [0.5 - 0.7]$) leads to better performance and a larger average number of particles. This behavior can also be seen in Figure \ref{fig_MSE_M}, where the MSE w.r.t. the number of particles is displayed for three different sets of thresholds. Note that a filter with a too relaxed set of thresholds ($[p_\ell - p_h]= [0.05 - 0.4]$) uses very few particles but obtains a poor performance, while a filter with the most stringent set of thresholds ($[p_\ell - p_h]= [0.5 - 0.9]$) consistently yields a low MSE, at the expense of using a larger number of particles.

The numerical results have been computed in a Matlab environment on a computer with an Intel Core i5 processor (2.7 GHz clock frequency) and 12 GB of RAM.

\begin{table*}
\setlength{\tabcolsep}{2pt}
\def\marginwidth{1.5mm}
\begin{center}
\begin{tabular}{|l@{\hspace{\marginwidth}}|c@{\hspace{\marginwidth}}|c@{\hspace{\marginwidth}}|c@{\hspace{\marginwidth}}|c@{\hspace{\marginwidth}}|c@{\hspace{\marginwidth}}|c@{\hspace{\marginwidth}}|c@{\hspace{\marginwidth}}|c@{\hspace{\marginwidth}}|c@{\hspace{\marginwidth}}|c@{\hspace{\marginwidth}}|}

\hline
\cline{2-4}
$[p_{l}-p_{h}]$  & {\bf Fixed $M=2^{15}$} & $[0.4 - 0.8 ]$ & $[0.35 - 0.7 ]$ & $[0.3 - 0.7 ]$ & $[0.25 - 0.65 ]$ & $[0.2 - 0.6 ]$ \\
\hline
\hline
 MSE  &1.5193 &1.5234 &1.5240 &1.5287 &3.7552 &4.6540 \\
\hline
 $\bar{M}$  &32768 &24951 &14840 &8729 &2197 &451 \\
\hline
 p-val  &0.5108 &0.5089 &0.4902 &0.4815 &0.4872 &0.4785 \\
\hline
 Hell. distance  &0.2312 &0.2355 &0.2493 &0.2462 &0.2476 &0.2521 \\
\hline
exec. time (s)  &6201 &5617 &3014 &1532 &131 &67 \\
\hline
time ratio & 1 &   1.10   & 2.1  &  4.05  & 47.43 & 92.36\\
\hline
\end{tabular}
\end{center}
\caption{Lorenz Model (Section \ref{model_lorenz}):  $\Delta=10^{-3}$, $T_{obs}=200\Delta$, $\sigma^2=0.5$. Algorithm details: $W=20$, $K=7$, $M_{\text{max}}=2^{15}$, $M_{\text{min}}=2^7$. MSE in the approximation of the posterior mean, averaged number of particles $\bar{M}$, averaged p-value, and averaged Hellinger distance.}
\label{table_decreasing_2}
\end{table*}

%%%%%%%%%%%%%%%%%%%%%%
\subsubsection{Multidimensional Observations} \label{ssLorenzMulti}
%%%%%%%%%%%%%%%%%%%%%%
Now we consider the case where we have observations also related to the second dimension of the hidden state. In particular, and following the notation of the previous experiment, we collect a sequence of bi-dimensional observations $\{ \Y_t \}_{t=1, 2, ...}$ with components
\begin{eqnarray}
Y_{1,t} = X_{1,400t} + V_{1,t}, \nonumber \\
Y_{2,t} = X_{2,400t} + V_{2,t}, \nonumber
\label{eqObservLorenzMulti}
\end{eqnarray}
where the observation noises $\{ V_{1,t} \}_{t=1, 2, ...}$ and $\{ V_{2,t} \}_{t=1, 2, ...}$ are two sequences of i.i.d. normal random variables with zero mean and variance $\sigma^2 = \frac{1}{2}$. Note that now the state is observed every $400$ discrete-time steps in order to make the system more difficult to be tracked.

The implemented algorithm is an extension of the unidimensional case, as suggested in Section \ref{ssMultidim}. In particular, we perform the assessment over the marginals $p(y_{i,t}|\y_{1:t-1})$, with $i=1,2$, and then, with both p-values, we adapt the number of particles as follows: if at least one of the marginals requires more particles, we increase the number of particles; if both marginals indicate no need for change of the number of particles, we keep it fixed; otherwise, we decrease the number.

Table \ref{table_lorenz_2d} shows the MSE in the approximation of the posterior mean, averaged number of particles $\bar{M}$, averaged p-value (over both dimensions), and the running time. Note that we can extract similar conclusions as in the case with scalar observations.

\begin{table*}
\setlength{\tabcolsep}{2pt}
\def\marginwidth{1.5mm}
\begin{center}
\begin{tabular}{|l@{\hspace{\marginwidth}}|c@{\hspace{\marginwidth}}|c@{\hspace{\marginwidth}}|c@{\hspace{\marginwidth}}|c@{\hspace{\marginwidth}}|c@{\hspace{\marginwidth}}|c@{\hspace{\marginwidth}}|c@{\hspace{\marginwidth}}|c@{\hspace{\marginwidth}}|c@{\hspace{\marginwidth}}|c@{\hspace{\marginwidth}}|c@{\hspace{\marginwidth}}|c@{\hspace{\marginwidth}}|}
\hline
\cline{2-4}
$[p_{l}-p_{h}]$  & $[0.4 - 0.8 ]$ & $[0.3 - 0.7 ]$ & $[0.3 - 0.65 ]$ & $[0.25 - 0.65 ]$ & $[0.2 - 0.6 ]$ & $[0.15 - 0.55 ]$ & $[0.1 - 0.5 ]$ \\
\hline
\hline
 MSE  &2.7151 &2.7131 &2.8351 &3.8862 &4.0814 &5.4015 &7.0323 \\
\hline
$\bar{M}$  &26175 &19652 &15788 &7761 &3858 &539 &203 \\
\hline
 p-val  &0.5020 &0.4953 &0.4858 &0.4906 &0.4914 &0.4820 &0.4869 \\
\hline
exec. time (s)  &2937.9851 &2120.0787 &1744.3426 &772.2125 &373.6780 &73.1735 &38.3487 \\\hline
\end{tabular}
\end{center}
\caption{Outputs of the particle filter with adaptive $M$ for the Lorenz model (Section \ref{model_lorenz}) with parameters $\Delta=10^{-3}$, $T_{obs}=400\Delta$, $\sigma^2=0.5$ and 2-dimensional observations. The algorithm parameters are chosen as $W=20$, $K=7$, $M_{\text{max}}=2^{15}$ and $M_{\text{min}}=2^7$. We display the MSE in the approximation of the posterior mean, the averaged number of particles $\bar{M}$, averaged p-value (over both dimensions), and the running time.}
\label{table_lorenz_2d}
\end{table*}

%%%%%%%%%%%%%%%%%%%%%%
\subsubsection{Discussion} \label{ssLorenzDiscussion}
\begin{figure}
\centering
\includegraphics[width=0.49\textwidth]{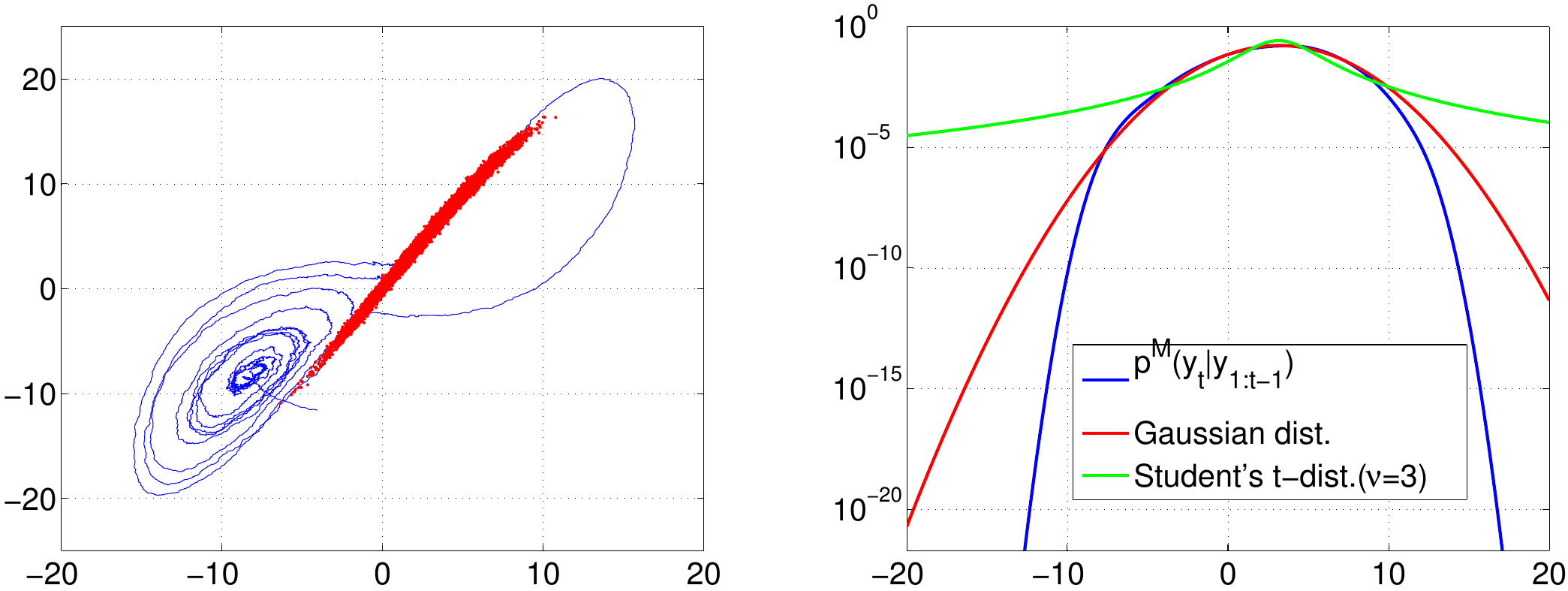}
\caption{Approximated observation predictive pdf $p^M(y_t|y_{1:t-1})$, Gaussian distribution, and Student's t-distribution ($\nu=3$) in log-scale for the stochastic Lorenz 63 example with $M=2^{14}$ particles. All distributions have the same mean and variance.} % t = 9601, n= 48
\label{fig_lorenz_hard}
\end{figure}

\begin{figure}
\centering
\includegraphics[width=0.49\textwidth]{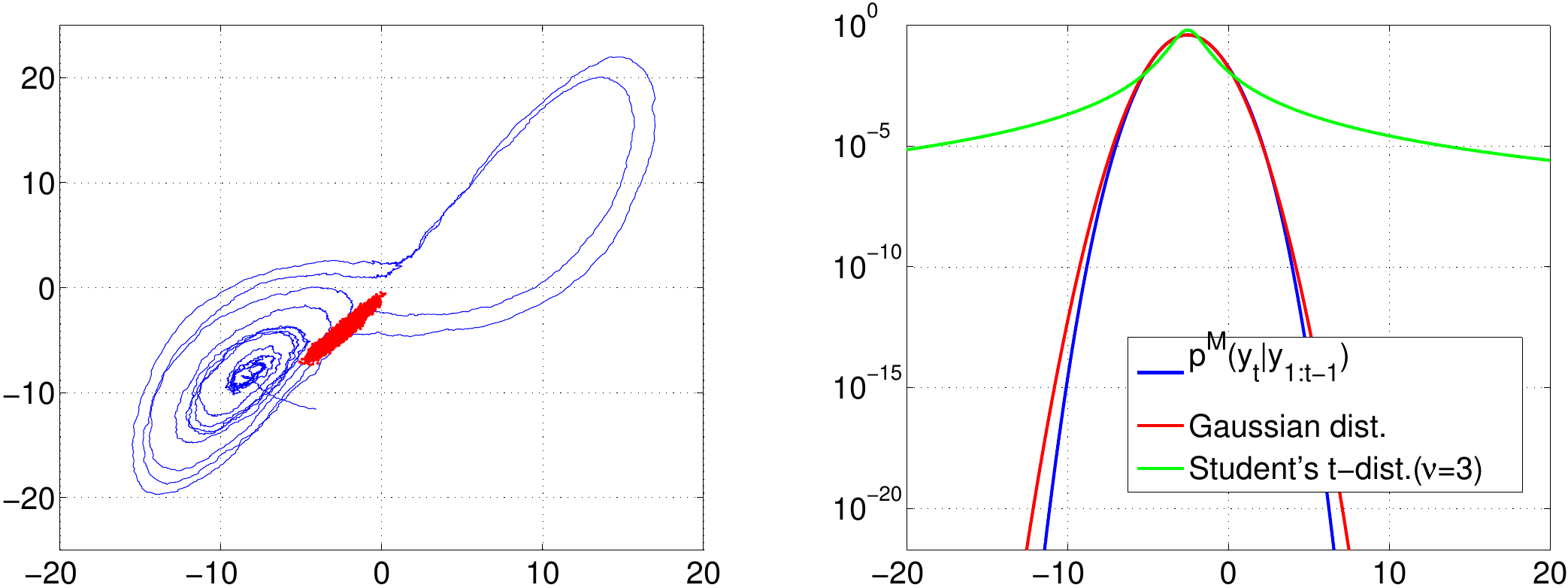}
\caption{Approximated observation predictive pdf $p^M(y_t|y_{1:t-1})$, Gaussian distribution, and Student's t-distribution ($\nu=3$) in log-scale for the stochastic Lorenz 63 example with $M=2^{14}$ particles. All distributions have the same mean and variance.} % t = 10201, n= 51
\label{fig_lorenz_easy}
\end{figure}

%%%%%%%%%%%%%%%%%%%%%%
The assumption \mfC~of Section {\ref{sec:convergence}} states that the tails of the pdf  $p(\y_t|\y_{1:t-1})$ should not be too heavy. Nevertheless, we have shown that the constraint is rather weak, since it is satisfied for all exponential-type distributions as well as for many heavy-tailed distributions. In practice, $p(\y_t|\y_{1:t-1})$ cannot be characterized for most models in a closed form. Here we show the particle approximation of the observation predictive pdf $p^M(\y_t|\y_{1:t-1})$ in the Lorenz 63 model at two different time steps. Figure {\ref{fig_lorenz_hard}} shows $p^M(\y_t|\y_{1:t-1})$ with $M=2^{14}$ particles in in log-scale at time $t = 9601$. The approximated pdf $p^M(\y_t|\y_{1:t-1})$ is compared with a Gaussian pdf and a Student's t-distribution (with $\nu=3$), all of them with the same mean and variance. Figure {\ref{fig_lorenz_easy}} shows the same distributions at a different time step ($t = 10201$). Note that  $p^M(\y_t|\y_{1:t-1})$ has very light tails at both time steps, and therefore, the assumption \mfC~ holds in both numerical examples.

%%%%%%%%%%%%%%%%%%%%%%
\subsection{Non-linear growth model with heavy-tailed observation noise} \label{ssGrowth}
%%%%%%%%%%%%%%%%%%%%%%

In this numerical example, we consider the problem of tracking a modified version of the non-linear growth model in \cite{Doucet00}. The state and observation equations are given by
\begin{eqnarray}
x_t&=&\frac{x_{t-1}}{2} + \frac{25x_{t-1}}{1+x_{t-1}^2} + 8\cos(\phi t) +  u_t,\\
y_t&=& \frac{x^2_t}{20}+ v_t,
\end{eqnarray}
where $\phi=0.4$ is a frequency parameter (in rad/s), $\{u_t\}_{t\ge 1}$ denotes a sequence of independent zero-mean univariate Gaussian r.v.'s with variance $\sigma_u^2=2$, and $\{v_t\}_{t\ge 1}$ is a sequence of independent Student's t-distributed r.v.'s with $\nu = 5$ degrees of freedom. The model is run for $t=1, 2, ...,T$, with $T=5,000$.

\begin{figure}
\centering
\includegraphics[width=0.5\textwidth]{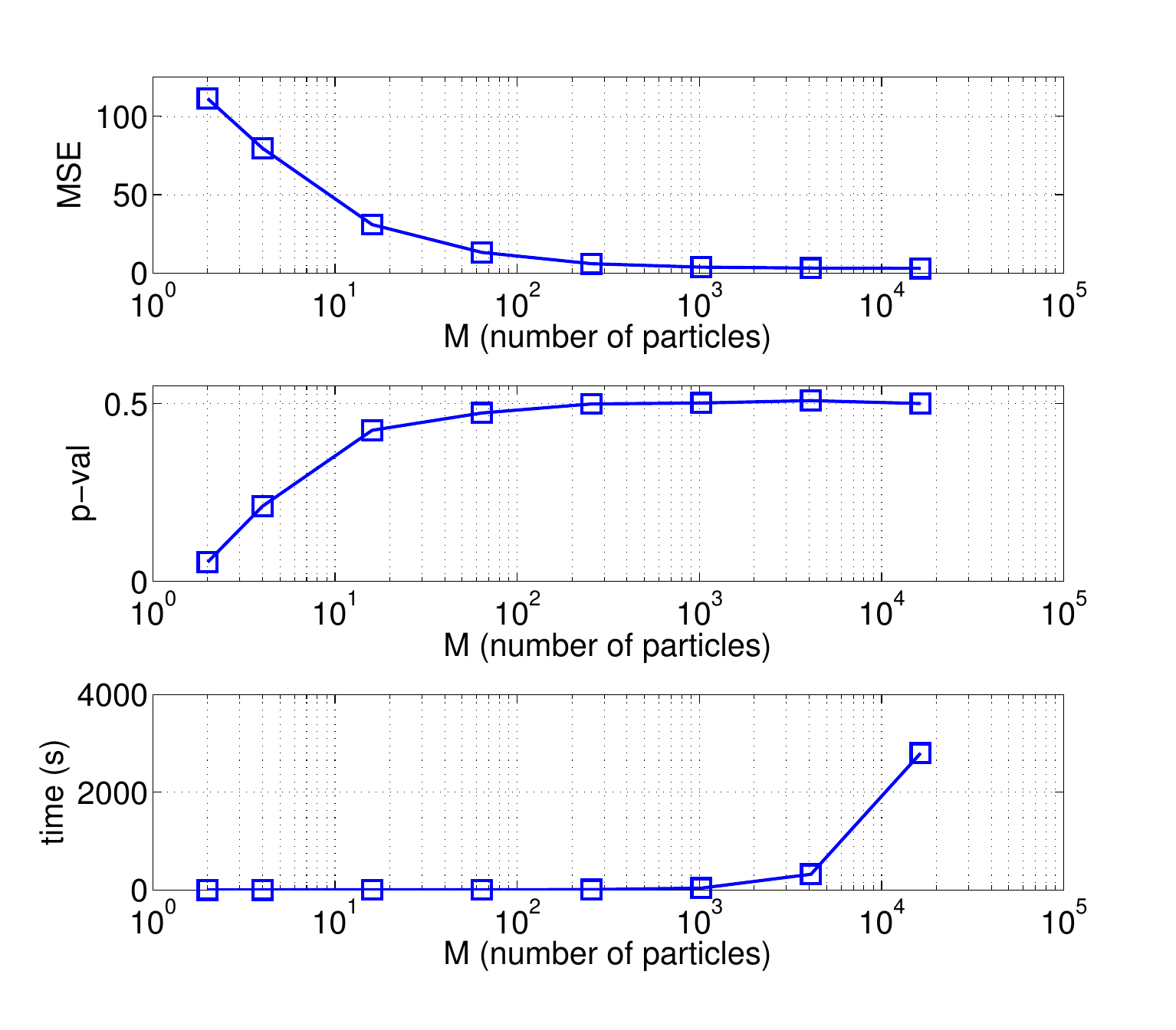}
\caption{BPF applied to a stochastic growth model with Student's t-distributed noise, and with fixed number of particles (Section \ref{ssGrowth}). MSE in the approximation of the posterior mean (top), averaged p-value (middle) and running time (bottom). The results are averaged over 50 independent simulations.}
\label{fig_sgt}
\end{figure}

First, we have run the standard BPF (with a fixed number of particles) for $M$ in the range between 2 and $2^{14}$. Figure \ref{fig_sgt}  shows, for each value of the fixed number of particles $M$, the MSE of the approximation of the posterior mean, the averaged p-value $p^*$ computed in the algorithm of Table \ref{table_algorithm}, and the running time. As expected, the MSE decreases with the number of particles, at the expense of increasing the computational complexity of the filter. Note also that, over a certain range of $M$ (namely, $M \ge 2^5$), the performance does not significantly improve. Finally, we see that in this example when the performance is poor, the p-value is very low (in average). This p-value is increased to $\approx 0.5$ when the performance of the filter improves.

Then, we have run the particle filter with adaptive number of particles in Table  \ref{table_algorithm}, with $K=5$ fictitious observations, window size $W=15$, p-value thresholds $[p_{l}-p_{h}] \in \{[0.4 - 0.68 ], [0.35 - 0.75 ], [0.3 - 0.7 ], [0.3 - 0.65 ], [0.25 - 0.65], [0.2 - 0.6]?\}$, initial number of particles $M_0=2^9$, maximum and minimum number of particles $M_{\text{max}}=2^{14}$ and $M_{\text{min}}=2^4$, respectively, $f_{\text{up}}(M_{n-1}) = 2 M_{n-1}$, and $f_{\text{down}}(M_{n-1}) = M_{n-1}/2$.

Table \ref{table_sgt} displays the MSE of the approximation of the posterior mean, the averaged number of particles, the average p-value, and the running time in seconds, for the different choices of $[p_{l}-p_{h}]$. The results are averaged over $50$ independent trials. Again, the pair of thresholds $[p_{l}-p_{h}]$ allows to operate at different complexity-performance regimes; decreasing the pair of parameters worsens the performance of the filter but enables a reduction in computational load.

\begin{table*}
\setlength{\tabcolsep}{2pt}
\def\marginwidth{1.5mm}
\begin{center}
\begin{tabular}{|l@{\hspace{\marginwidth}}|c@{\hspace{\marginwidth}}|c@{\hspace{\marginwidth}}|c@{\hspace{\marginwidth}}|c@{\hspace{\marginwidth}}|c@{\hspace{\marginwidth}}|c@{\hspace{\marginwidth}}|c@{\hspace{\marginwidth}}|c@{\hspace{\marginwidth}}|c@{\hspace{\marginwidth}}|c@{\hspace{\marginwidth}}|c@{\hspace{\marginwidth}}|c@{\hspace{\marginwidth}}|}
\hline
\cline{2-4}
$[p_{l}-p_{h}]$  & $[0.4 - 0.8 ]$ & $[0.35 - 0.75 ]$ & $[0.3 - 0.7 ]$ & $[0.3 - 0.65 ]$ & $[0.25 - 0.65 ]$ & $[0.2 - 0.6 ]$ \\
\hline
\hline
 MSE  &2.8707 &3.4945 &4.7687  &9.0465 &10.5826 &17.6967 \\
\hline
 $\bar{M}$  &9739 &7478 &6251 &3168 &2087 &232 \\
\hline
 p-val  &0.4976 &0.4950 &0.4893 &0.4837 &0.4730 &0.4604 \\
\hline
exec. time (s)  &3613  &2515 &1427 &561 &234 &21 \\
\hline
\end{tabular}
\end{center}
\caption{Output of the algorithm in Table \ref{table_algorithm} for a stochastic growth model with Student's t-distributed observation noise, with adaptive $M$ (Section \ref{ssGrowth}). The algorithm parameters are chosen as $W=15$, $K=1$, $M_{\text{max}}=2^{14}$, $M_{\text{min}}=2^6$. We display the MSE in the approximation of the posterior mean, the average number of particles $\bar{M}$, the average p-value, and the running time.}
\label{table_sgt}
\end{table*}

%%%%%%%%%%%%%%%%%%%%%%
%%%%%%%%%%%%%%%%%%%%%%
%%%%%%%%%%%%%%%%%%%%%%
\section{Conclusions}
\label{sec_conclusions}
%%%%%%%%%%%%%%%%%%%%%%
%%%%%%%%%%%%%%%%%%%%%%
%%%%%%%%%%%%%%%%%%%%%%

In practice, the number of particles needed in a particle filter is usually determined in an ad hoc manner. Furthermore, this number is typically kept constant throughout tracking. In this paper, we have proposed a methodology for the online determination of the number of particles needed by the filter. The approach is based on assessing the convergence of the predictive distribution of the observations online. First we have proved, under standard assumptions, a novel convergence result on the approximation of this distribution. Then, we have proposed a method for adapting the number of particles based on the online assessment of the filter convergence. We have illustrated the performance of the suggested algorithm by computer simulations. The proposed procedure is simple but not unique. Namely, with the proposed methodology one can develop a range of algorithms for adapting the number of particles. Furthermore, while the analysis and examples have been presented for the standard bootstrap particle filter for simplicity and clarity, it is straightforward to extend it to more sophisticated algorithms using adaptive proposals \cite{cornebise2008adaptive} or parallelization schemes \cite{whiteley2016role,paige2014asynchronous}.

%%%%%%%%%%%%%%%%%%%%%%%%%%%%%%%%%%%
\appendices

%%%%%%%%%%%%%%%%%%%%%%
%%%%%%%%%%%%%%%%%%%%%%
%%%%%%%%%%%%%%%%%%%%%%
\section{Proof of Theorem \ref{th1}}
\label{apTh1}
%%%%%%%%%%%%%%%%%%%%%%
%%%%%%%%%%%%%%%%%%%%%%
%%%%%%%%%%%%%%%%%%%%%%

Recall that the likelihood of $\X_t=\x_t$ given the observation $\Y_t=\y_t$ is denoted $g_t^{\y_t}(\x_t)$, i.e., $g_t^{\y_t}(\x_t)=p(\y_t|\x_t)$. For the sake of notational accuracy, we introduce the Markov transition kernel $\tau_t(d\x_t|\x_{t-1})$ that determines the dynamics of the state process. This kernel is connected to the notation in Section \ref{sPF} by $\tau_t(d\x_t|\x_{t-1}) = p(\x_t|\x_{t-1})d\x_t$. However, all the results in this appendix (including Theorem \ref{th1}) are proved for the general case in which $\tau_t$ does not necessarily have a density w.r.t. the Lebesgue measure. For notational coherence, we denote $\tau_0(d\x_0)=p(\x_0)d\x_0$.

The same as in Section \ref{sPF}, the integral of a function $f:\mZ\rw\Real$ w.r.t. a measure $\alpha$ on the measurable space $(\mB(\mZ),\mZ)$ is denoted $(f,\alpha)$ and the absolute supremum of $f$ is written $\|f\|_\infty=\sup_{\z\in\mZ}|f(\z)|$. The class of bounded real functions over the set $\mZ$ is denoted $B(\mZ) = \{ f:\mZ\rw\Real: \|f\|_\infty<\infty\}$. For $p \ge 1$, the $L_p$ norm of a r.v. $Z$ with associated probability measure $\gamma(dz)$ is denoted 
$$
\| Z \|_p := E\left[  |Z|^p \right]^{\frac{1}{p}} = \left(
	\int |z|^p \gamma(dz)
\right)^\frac{1}{p},
$$ 
where $E[\cdot]$ denotes expectation.

We start introducing some auxiliary results on the convergence of the approximate measure $\xi_t^M$ (to be precise, on the convergence of approximate integrals of the form $(f_t, \xi_t^M)$, where $f_t \in B(\mX)$). This leads to the core result, which is the uniform convergence of $p_t^M(\y_t) \rw p_t(\y_t)$ on a sequence of compact sets. The proof of Theorem \ref{th1} follows readily from the latter result.

The analysis in this Appendix draws from methods developed in \cite{crisan2014particle} for the estimation of the filter pdf $p(\x_t|\y_{1:t})$ using kernel functions, which herein are suitably adapted to the problem of approximating the predictive density $p_t(\y_t)$.

%%%
\begin{Lema} \label{lmXi-1}
Assume that the sequence $\y_{1:T}$, for $T<\infty$, is arbitrary but fixed, and, for each $t=1, 2, ..., T$, $g_t^{\y_t} \in B(\mX)$ and $g_t^{\y_t}>0$. Then, there exist constants $c_t<\infty$, $t=1, 2, ..., T$, independent of $M$ such that
\begin{equation}
\| (f,\xi_t^M) - (f,\xi_t) \|_p \le \frac{
	c_t\|f\|_\infty
}{
	\sqrt{M}
}, \quad t =  1, 2, 3...
\nonumber
\end{equation}
for every $f \in B(\mX)$. 
\end{Lema}
%%%
\noindent\textbf{\textit{Proof}}: This is a particular case of \cite[Lemma 1]{Miguez13b}. \qed
%%%

%%%
\begin{Nota}{(\textbf{The `standard setting'})}
Most of the results proved in this Appendix rely on Lemma \ref{lmXi-1} and, therefore, are only true under the basic assumptions of that Lemma. For conciseness, we will say that a result holds {\em within the standard setting} when we assume that {\em the sequence $\y_{1:T}$, for $T<\infty$, is arbitrary but fixed, and, for each $t=1, 2, ..., T$, $g_t^{\y_t} \in B(\mX)$ and $g_t^{\y_t}>0$}, and so Lemma \ref{lmXi-1} can be applied.
\end{Nota}
%%%

For each pair of natural numbers $M$ and $d$, we introduce a family of function-valued r.v.'s, denoted as $\sfF_t^M(d)$ and explicitly defined below.
%%%
\begin{Definicion} \label{defFt}
A function $\sff_t^M : \Real^d \rw \Real$ belongs to the family $\sfF_t^M(d)$ if, and only if, for every $\y \in \Real^d$ we can express $\sff_t^M(\y)$ as
\begin{equation}
\sff^M(\y) = (a^\y,\xi_t^M) - (a^\y,\xi_t),
\nonumber
\end{equation}
where $a^\y(\x)=a(\x,\y)$ is a bounded function $\mX \times \Real^d \rw \Real$ with bounded derivatives of order up to $d$ w.r.t. the variable $\y$, specifically,
\begin{equation}
\sup_{(\x,\y) \in \mX \times \Real^d} | a(\x,\y) | < \infty 
\quad \mbox{and} \quad 
\sup_{(\x,\y) \in \mX \times \Real^d} |D^{\bf 1} a(\x,\y)| <\infty,
\nonumber 
\end{equation}
where the partial derivative operator acts on $\y$, i.e., $D^{\bf 1}a(\x,\y) = \frac{ \partial^d a }{ \partial y_1 \cdots \partial y_d}(\x,\y)$. 
\end{Definicion}
%%%
We use the notation $\| a \|_\infty \dfn \sup_{(\x,\y) \in \mX \times \Real^d} | a(\x,\y) |$ and $\| D^{\bf 1} a \|_\infty \dfn \sup_{(\x,\y) \in \mX \times \Real^d} |D^{\bf 1} a(\x,\y)|$. It is apparent that $a^\y \in B(\mX)$, hence the estimate $(a^\y,\xi_t^M)$ converges to $(a^\y,\xi_t)$ when $M\rw\infty$ as given by Lemma \ref{lmXi-1}. Also note that $\sff_t^M$ is a function-valued r.v. measurable w.r.t the $\sigma$-algebra generated by $\{\bar\x_t^{(m)}\}_{m=1,...,M}$. The following lemma provides upper bounds on the moments of the members of $\sfF_t^M(d)$.

%%%
\begin{Lema} \label{lmPt-1}
Within the standard setting, for every $\sff^M_t \in \sfF_t^M(d)$ and every $p \ge 1$ there exists a constant $\bar c_t<\infty$ independent of $M$ and $\y$ such that
\begin{eqnarray}
E\left[ \left| \sff_t^M(\y) \right|^p \right] \leq \frac{ \bar c_t^p }{ M^{\frac{p}{2}} }.
\label{eqkk0}
\end{eqnarray}
\end{Lema}
\noindent{\textbf{Proof:}} From the definition of the family $\sfF_t^M(d)$, we can write for every $\y \in \Real^d$, 
\begin{equation}
\| \sff_t^M(\y) \|_p = \| (a^\y,\xi_t^M) - (a^\y,\xi_t) \|_p \label{eqPrf0}
\end{equation}
for some $a^\y \in B(\mX)$, with an upper bound $\| a \|_\infty < \infty$ uniform over $\y \in \Real^d$. However, \eqref{eqPrf0} together with Lemma \ref{lmXi-1} yields
\begin{equation}
\| \sff_t^M(\y) \|_p \le \frac{\bar c_t}{\sqrt{M}}
\label{eqPrf1}
\end{equation}
where $\bar c_t = c_t \| a \|_\infty < \infty$ is independent of $M$ and $\y$. If we raise both sides of \eqref{eqPrf1} to power $p$, then we obtain the desired result of \eqref{eqkk0}. \hfill \qed
%%%

%%%
\begin{Lema}\label{lmZM}
\label{lema_sequence}
Let $\{ \theta^M \}_{M \geq1 }$ be a sequence of non-negative r.v.'s such that, for every $p \ge 4$,
\begin{equation}
E\left[ \left( \theta^M \right)^p \right] \leq \frac{c}{M^{\frac{p}{2}-\nu}}
\label{eqkk2}
\end{equation}
where $c<\infty$ and $0 \le \nu < 1$ are constants independent of $M$. Then, for every $\epsilon \in (0,\frac{1}{2})$ there exists an a.s. finite r.v. $U^\epsilon$ independent of $M$ such that
\begin{equation}
\theta^M \leq \frac{U^\epsilon}{M^{\frac{1}{2}-\epsilon}}.
\nonumber
\end{equation}
\end{Lema}
%%%
\noindent{\textbf{Proof}}:
Let us choose an arbitrary constant $\psi \in (\nu, 1)$ and define the r.v. $U^{\psi,p} = \sum_{M=1}^\infty M^{\frac{p}{2}-1-\psi}(\theta^M)^p$. If \eqref{eqkk2} holds, then the expectation $E[U^{\psi,p}]$ is finite, as we prove in the sequel. Indeed, from Fatou's lemma,
\begin{eqnarray}
E\left[
	U^{\psi,p} 
\right] &\leq& \sum_{M=1}^\infty M^{\frac{p}{2}-1-\psi} E\left[
	(\theta^M)^p
\right] \label{eqkk3}\\
&\leq& c \sum_{M=1}^\infty M^{\nu-\psi-1},
\label{eqkk4}
\end{eqnarray}
where \eqref{eqkk4} follows from substituting \eqref{eqkk2} into \eqref{eqkk3}. Since we have chosen $\psi \in (\nu,1)$, then it follows that $-1 < \nu-\psi < 0$ and $\nu-\psi-1 < -1$, which ensures that $\sum_{M=1}^\infty M^{\nu-\psi-1} < \infty$ and, therefore, $E\left[
	U^{\psi,p} 
\right] < \infty$. Since $E\left[
	U^{\psi,p} 
\right] < \infty$, then $U^{\psi,p} < \infty$ a.s.

For any given value of $M$, it is apparent from the definition of $U^{\psi,p}$ that 
\begin{equation}
 M^{\frac{p}{2}-1-\psi} (\theta^M)^p \leq U^{\psi,p}
 \nonumber
\end{equation}
and, as a consequence, 
\begin{equation}
\theta^M \leq \frac{
	(U^{\psi,p})^{\frac{1}{p}}
}{
	M^{\frac{1}{2} - \frac{1+\psi}{p}}
} = \frac{U^\epsilon}{M^{\frac{1}{2}-\epsilon}}
\label{eqkk5}
\end{equation}
where the equality in \eqref{eqkk5} follows from defining $\epsilon \dfn \frac{1+\psi}{p}$ and $U^\epsilon \dfn (U^{\psi,p})^{\frac{1}{p}}$. Since $\psi < 1$, it is sufficient to choose $p\ge4$ to ensure that $\epsilon = \frac{1+\psi}{p} < \frac{1}{2}$. Also, since $p$ can actually be chosen as large as we wish, it follows that \eqref{eqkk5} holds for $\epsilon>0$ as small as needed. \hfill \qed
%%%

%%%
% Explicit random error rates for \xi_t^M
%%%
%%%
\begin{Lema} \label{lmXi-2}
%Assume that the sequence $\y_{0:T}$, for $T<\infty$, is arbitrary but fixed, and, for $t=1, 2, ..., T$, $g_t^{\y_t} \in B(\mX)$ and $g_t^{\y_t}>0$. Then, 
Within the standard setting, for every $0 < \epsilon < \frac{1}{2}$ (arbitrarily small) there exist a.s. finite r.v.'s $U_t^\epsilon<\infty$, {$t=1, 2, ..., T$}, independent of $M$ such that
\begin{equation}
| (f,\xi_t^M) - (f,\xi_t) | \le \frac{
	U_t^\epsilon
}{
	M^{\frac{1}{2}-\epsilon}
}, \quad {t = 1, 2, 3, ...}
\label{eqkk6}
\end{equation}
for every $f \in B(\mX)$. 
\end{Lema}
%%%
\noindent\textbf{\textit{Proof}}: From Lemma \ref{lmXi-1}, for each $t=1, ..., T$, there is a constant $c_t$ independent of $M$ such that 
$$
E\left[
	| (f,\xi_t^M) - (f,\xi_t) |^p 
\right] \le \frac{c_t^p\|f\|_\infty^p}{M^{\frac{p}{2}}}
$$ 
for any $f \in B(\mX)$. Therefore, we can apply Lemma \ref{lmZM} with $c=c_t^p\|f\|_\infty^p$ and $\nu = 0$, to obtain the desired inequality \eqref{eqkk6}. \hfill \qed
%%%

For the statement of the next result, we need to recall the definition of the sequence of hypercubes 
$$
C_M := \left[ -\frac{M^\frac{\beta}{d_y p}}{2}, +\frac{M^\frac{\beta}{d_y p}}{2} \right] \times \cdots \times \left[ -\frac{M^\frac{\beta}{d_y p}}{2}, +\frac{M^\frac{\beta}{d_y p}}{2} \right] \subset \Real^{d_y}
$$ 
in assumption \mfC, where $p \ge 4$ and $0<\beta<1$ are constants w.r.t. $M$.

%%%
\begin{Lema} \label{lmUnif}
Within the standard setting, for any $0 < \varepsilon < \frac{1}{2}$, every $\sff_t^M \in \sfF_t^M(d_y)$ and each $t=1, 2, ..., T$ there exists an a.s. finite r.v. $V_t^\varepsilon$ independent of $M$ such that
\begin{eqnarray}
\sup_{\y \in C_M} | \sff_t^M(\y) | \leq \frac{V_t^\varepsilon}{M^{\frac{1}{2}-\varepsilon}}.
\label{eqFtStatement}
\end{eqnarray}
In particular,
\begin{equation}
\lim_{M \rightarrow \infty} \sup_{\y \in C_M} | \sff_t^M(\y) |=0 \quad \mbox{a.s.}
\nonumber
\end{equation}
\end{Lema}
%%%
\noindent\textbf{\textit{Proof}}:
Let $b_M=\frac{1}{2}M^\frac{\beta}{d_y p}$, in such a way that the hypercube $C_M$ can be written as $C_M=[-b_M, +b_M]^{d_y}\subset\Real^{d_y}$. We prove that the inequality \eqref{eqFtStatement} holds by induction in the dimension $d_y$.

We start with the case $d_y=1$, hence $\mY \subseteq \Real$ and the observations $\y_t=y_t \in \mY$ are scalars. From Definition \ref{defFt}, any $\sff_t^M \in \sfF_t^M(1)$ is differentiable in every interval $C_M$, hence we can apply the fundamental theorem of calculus (FTC) to express $\sff_t^M(y)$, for $-b_M \le y \le b_M$, as
\begin{equation}
\sff_t^M(y) = \sff_t^M(0) + \int_{-b_M}^y \frac{d\sff_t^M}{dy}(z)dz -  \int_{-b_M}^0 \frac{d\sff_t^M}{dy}(z)dz.  
\nonumber
\end{equation}
As a consequence, we obtain a simple upper bound for the magnitude of $\sff_t^M(y)$, namely
\begin{equation}
\sup_{y \in C_M} |\sff_t^M(y)| \le |\sff_t^M(0)| + 2A^M, \label{eqSupsff}
\end{equation}
where 
\begin{equation}
A^M = \int_{-b_M}^{b_M} \left| \frac{d\sff_t^M}{dy}(z) \right |dz.
\end{equation}

In order to find an upper bound for the term $A^M$, we apply Jensen's inequality, which yields, for $p \ge 1$, 
\begin{equation}
\left(
        \frac{1}{2 b_M} A^M
\right)^p \le
\frac{1}{2 b_M} \int_{-b_M}^{b_M}  \left| \frac{d\sff_t^M}{dy}(z) \right |^p dz \label{eqJensenIneq}
\end{equation}
and the inequality \eqref{eqJensenIneq} above readily leads to
\begin{equation}
\left(
        A^M
\right)^p \le 2^{p-1} b_M^{p-1} \int_{-b_M}^{b_M}  \left| \frac{d\sff_t^M}{dy}(z) \right |^p dz.
\label{eqBoundBigA}
\end{equation}
However, since $\sff_t^M \in \sfF_t^M(1)$, there exists some function $a(x,y)$ such that
\begin{equation}
\frac{d\sff_t^M}{dy}(y) = \left(
	\frac{\partial a}{\partial y}(x,y),\xi_t^M
\right) - \left(
	\frac{\partial a}{\partial y}(x,y),\xi_t
\right),
\nonumber
%\label{eqAlmendruco}
\end{equation}
with $\left\| \frac{\partial a}{\partial y} \right\|_\infty = \sup_{(x,y)\in\mX\times\mY} \left|\frac{\partial a}{\partial y}(x,y) \right| < \infty$. Since $\frac{\partial a^y}{\partial y} = \frac{\partial a}{\partial y}(\cdot,y) \in B(\mX)$, we can apply Lemma \ref{lmPt-1} to arrive at
\begin{eqnarray}
E\left[
	\left| 
		\frac{d\sff_t^M}{dy}(y)
	\right|^p
\right] &=& E\left[
	\left|
		\left(
			\frac{\partial a^y}{\partial y},\xi_t^M
		\right) - \left(
			\frac{\partial a^y}{\partial y},\xi_t
		\right)
	\right|^p
\right] \nonumber\\
&\le& \frac{
        \bar c_t^p
}{
        M^\frac{p}{2}
},
\label{eqSmallBoundAk}
\end{eqnarray}
where the constant $\bar c_t^p \propto \left\| \frac{\partial a}{\partial y} \right\|_\infty^p$ is independent of $M$ and $y$. We can combine \eqref{eqSmallBoundAk} and \eqref{eqBoundBigA} to arrive at
\begin{equation}
E\left[
        (A^M)^p
\right] \le \frac{
        2^p b_M^p \bar c_t^p 
}{
        M^\frac{p}{2}
} = \frac{
        \bar c_t^p 
}{
        M^{\frac{p}{2}-\beta}
}, \nonumber
\end{equation}
where the equality follows from the relationship $b_M = \frac{1}{2}M^\frac{\beta}{p}$. 

If we now apply Lemma \ref{lmZM} with $\theta^M = A^M$, $p \ge 4$, $\nu=\beta$ and $c=\bar c_t^p$, then we obtain a constant $\varepsilon_1 \in \left( \frac{1+\beta}{p}, \frac{1}{2} \right)$ (see \eqref{eqkk5}) and a non-negative and a.s. finite random variable $V^{A,\varepsilon_1}$, both of them independent of $M$ and $y$, such that
\begin{equation}
A^M \le \frac{
        V^{A,\varepsilon_1}
}{
        M^{\frac{1}{2}-\varepsilon_1}
}. \label{eqRate_part1}
\end{equation} 

Moreover, from Lemma \ref{lmXi-2} we readily obtain the inequality
%Lemma \ref{lmPt-1} also yields 
%\begin{eqnarray}
%E\left[
%        \left|
%                \sff_t^M(0)
%        \right|^p
%\right] \le \frac{
%        \tilde c_t^p
%}{
%        M^{\frac{p}{2}}
%}, \nonumber
%\end{eqnarray}
%where $\tilde c_t<\infty$ is a constant independent of $M$ and $y$. Therefore, we can apply Lemma \ref{lmZM} again, with $\theta^M = | \sff_t^M(0) |$, $p \ge 4$, $\nu=0$ and $c=\tilde c_t^p$ to obtain the inequality
\begin{equation}
\left|
        \sff_t^M(0)
\right| \le \frac{
        V^{0,\varepsilon_2}
}{
        M^{\frac{1}{2}-\varepsilon_2}
}, \label{eqRate_part2}
\end{equation}
where $\varepsilon_2 \in \left(0,\frac{1}{2}\right)$ is a constant and $V^{0,\varepsilon_2}$ is a non-negative and a.s. finite r.v., both of them independent of $M$. 

If we choose $\varepsilon=\varepsilon_1=\varepsilon_2 \in \left( \frac{1+\beta}{p}, \frac{1}{2} \right)$ and define $V_t^\varepsilon = 2V^{A,\varepsilon_1} + V^{0,\varepsilon_2}$, then the combination of Eqs. \eqref{eqSupsff}, \eqref{eqRate_part1} and \eqref{eqRate_part2} yields 
\begin{equation}
\sup_{\y \in C_M} \left|
        \sff_t^M(y)
\right| \le \frac{V_t^\varepsilon}{M^{\frac{1}{2}-\varepsilon}},
\nonumber
\end{equation}
where $V_t^\varepsilon$ is a.s. finite. Note that $V_t^\varepsilon$ and $\varepsilon$ are independent of $M$ and $y$. Moreover, we can choose $p$ as large as we wish and $\beta>0$ as small as needed, hence we can effectively select $\varepsilon \in (0,\frac{1}{2})$. This completes the analysis for $d_y=1$.

Next, we assume that the inequality \eqref{eqFtStatement} holds for $d_y=d-1 > 1$ and show that, in such case, it also holds for $d_y=d$.

Let us initially analyze $\sff_t^M(\y)$ for $\y \in \left[0,M^{\frac{\beta}{dp}}\right]^d$ (i.e., $\y=y_{1:d}$ with $y_i > 0$ for every $i=1, ..., d$). Using the FTC we obtain
\begin{eqnarray}
\sff_t^M(y_{1:d}) &=& \sff_t^M(y_{1:d-1}, 0) + \nonumber\\
&& \int_0^{y_d} D^{\alpha_1} \sff_t^M(y_{1:d-1}, z_d ) dz_d, \nonumber\\
\label{eqX0}
\end{eqnarray}
where $\alpha_1=(0, ..., 0, 1)$. The function in the integral of the right hand side (rhs) of \eqref{eqX0} can be expanded, using the FTC again, as 
\begin{eqnarray}
D^{\alpha_1} \sff_t^M(y_{1:d-1}, z_d ) &=& \nonumber \\
D^{\alpha_1} \sff_t^M(y_{1:d-2}, 0, z_d ) &&\nonumber\\
+ \int_0^{y_{d-1}} D^{\alpha_2} \sff_t^M(y_{1:d-2}, z_{d-1},z_d) dz_d dz_{d-1},&&
\label{eqX1}
\end{eqnarray}
where $\alpha_2 = (0, ..., 0, 1, 1)$. Substituting \eqref{eqX1} into \eqref{eqX0} yields
\begin{eqnarray}
\sff_t^M(y_{1:d}) &=& \nonumber\\
\sff_t^M(y_{1:d-1}, 0) + &&\nonumber\\
\int_0^{y_d} D^{\alpha_1} \sff_t^M(y_{1:d-2}, 0, z_d ) dz_d + &&\nonumber\\
\int_0^{y_d} \int_0^{y_{d-1}} D^{\alpha_2} \sff_t^M(y_{1:d-2}, z_{d-1},z_d) dz_d dz_{d-1}.&&
\label{eqX2}
\end{eqnarray}
By successively applying the FTC $d-2$ more times, \eqref{eqX2} becomes
\begin{eqnarray}
\sff_t^M(y_{1:d}) &=& \sum_{i=0}^{d-1} \tilde \sff_{t,i}^M(\y_{(d-i)}) + \nonumber\\
&& \int_0^{y_d} \cdots \int_0^{y_1} D^{\bf 1} \sff_t^M(z_{1:d}) dz_d \cdots dz_1,
\label{eqX3}
\end{eqnarray}
where $\y_{(d-i)} = (y_{1:d-i-1}, y_{d-i+1:d}) \in \Real^{d-1}$,
\begin{eqnarray}
\tilde \sff_{t,0}^M(\y_{(d)}) &\dfn& \sff_t^M(y_{1:d-1}, 0), \label{eqX4} \\
\tilde \sff_{t,i}^M(\y_{(d-i)}) &\dfn& \int_0^{y_d} \cdots \int_0^{y_{d-i+1}} \nonumber\\
&&D^{\alpha_i} \sff_t^M(y_{1:d-i-1}, 0, z_{d-i+1:d}) dz_{d-i+1:d}. \nonumber\\
&& \label{eqX5}
\end{eqnarray}
and $\alpha_i=(\overbrace{0, ..., 0}^{d-i}, \overbrace{1, ..., 1}^{i})$. From Eq. \eqref{eqX3} we readily obtain the bound
\begin{eqnarray}
\left| 
	\sff_t^M(y_{1:d}) 
\right| &\le& \sum_{i=0}^{d-1} \left| 
	\tilde \sff_{t,i}^M(\y_{(d-i)}) 
\right| + \nonumber\\
&& \int_0^{b_M} \cdots \int_0^{b_M} \left|
	D^{\bf 1}_y \sff_t^M(z_{1:d}) 
\right| dz_d \cdots dz_1, \nonumber\\
\label{eqX6}
\end{eqnarray}
that holds for the case $0 \le y_i \le b_M=\frac{1}{2}M^{\frac{\beta}{dp}}$, $i=1, 2, ..., d$.

By inspecting \eqref{eqX4} and \eqref{eqX5} we realize that if $\sff_t^M \in \sfF_t^M(d)$, then $\tilde \sff_{t,i}^M \in \sfF_t^M(d-1)$ for $i=0, 1, ..., d-1$. Therefore, from the induction hypothesis (and the fact that $M^{\frac{\beta}{(d-1)p}} \ge M^{\frac{\beta}{dp}}$) we deduce that, for any $\varepsilon \in (0,\frac{1}{2})$ there exist a.s. finite r.v.'s $\tilde V_i^\varepsilon$, $i=0,1, ..., d-1$, such that
\begin{equation}
\sup_{\y \in \left[ 0, \frac{1}{2}M^{\frac{\beta}{dp}} \right]^{d-1}}
\left|
	\tilde \sff_{t,i}^M(\y)
\right| \le \frac{\tilde V_i^\varepsilon}{M^{\frac{1}{2}-\varepsilon}}.
\label{eqX7}
\end{equation} 

As for the $d$-dimensional integral on the rhs of \eqref{eqX6}, we can find a suitable upper bound by the same procedure as in the base case, as shown below. Let $\z = z_{1:d}$ and denote, for $d>1$,
$$
A_d^M = \int_{0}^{b_M}\cdots\int_{0}^{b_M} \left|
        D^{\bf 1} \sff_t^M(\z)d\z
\right| d\z.
$$
An application of Jensen's inequality yields, for $p \ge 1$, 
\begin{equation}
\left(
        \frac{1}{b_M^d} A_d^M
\right)^p \le \frac{1}{b_M^d} \int_{0}^{b_M}\cdots\int_{0}^{b_M} \left|
        D^{\bf 1} \sff_t^M(\z)
\right|^p d\z, \nonumber
\end{equation}
which leads to
\begin{equation}
\left(
        A_d^M
\right)^p \le b_M^{d(p-1)} \int_{0}^{b_M}\cdots\int_{0}^{b_M} \left|
        D^{\bf 1} \sff_t^M(\z) 
\right|^p d\z. %\nonumber\\
\label{eqBoundBigA_d}
\end{equation}
Since, from Lemma \ref{lmPt-1},
\begin{equation}
E\left[
        \left|
                D^{\bf 1} \sff_t^M(\z)
        \right|^p
\right] \le \frac{
        \bar c_t^p
}{
        M^\frac{p}{2}
},
\label{eqSmallBoundAk_d}
\end{equation}
independently of $\z$, we can combine \eqref{eqSmallBoundAk_d} and \eqref{eqBoundBigA_d} to arrive at
\begin{equation}
E\left[
        (A^M)^p
\right] \le \frac{
        b_M^{d p} \bar c_t^p 
}{
        M^\frac{p}{2}
} = \frac{1}{2^{dp}} \times \frac{
        \bar c_t^p 
}{
        M^{\frac{p}{2}-\beta}
}, \nonumber
\end{equation}
where the equality follows from the relationship $b_M = \frac{1}{2}M^\frac{\beta}{d_y p}$. If we now apply Lemma \ref{lmZM} with $\theta^M = A^M$, $p \ge 4$, $\nu=\beta$ and $c=2^{-dp} \bar c_t^p$, then we deduce that for any constant $\varepsilon \in \left( \frac{1+\beta}{p}, \frac{1}{2} \right)$ (see \eqref{eqkk5}) there exists a non-negative and a.s. finite r.v. $V_A^{\varepsilon}$ (with both $\epsilon$ and $V_A^\varepsilon$ independent of $M$) such that
\begin{equation}
A_d^M \le \frac{
        V_A^{\varepsilon}
}{
        M^{\frac{1}{2}-\varepsilon}
}. \label{eqRate_part1_d}
\end{equation} 

Taking the inequalities \eqref{eqX6}, \eqref{eqX7} and \eqref{eqRate_part1_d} together, we arrive at
\begin{equation}
\sup_{\y \in \left[ 0, \frac{1}{2}M^{\frac{\beta}{dp}} \right]^d} | \sff_t^M(\y) | \le \frac{ V_0^\varepsilon }{ M^{\frac{1}{2}-\varepsilon} }
\label{eqX8}
\end{equation}
that holds for any constant $\varepsilon \in \left( \frac{1+\beta}{p}, \frac{1}{2} \right)$ and the a.s. finite r.v. $V_0^\varepsilon = V_A^\varepsilon + \sum_{i=0}^{d-1} \tilde V_i^\varepsilon$. Since we can select $p$ as large as we need, then we can effectively choose $\varepsilon \in (0,\frac{1}{2})$.
 
 To conclude the proof, we need to extend the bound in \eqref{eqX8} to the complete hypercube $C_M \subset \Real^d$. This is relatively straightforward. Assume, for example, that we have $\y = y_{1:d}$ such that $y_i \in [0,b_M]$ for $i=1, ...,d-1$ but $y_d \in [-b_M,0)$. Then we can consider the function $\breve \sff_t^M(y_{1:d-1},y_d) \dfn \sff_t^M(y_{1:d-1},-y_d) \in \sfF_t^M(d)$ and repeat the analysis to obtain the same type of bound as in \eqref{eqX8}. Indeed, we can classify every $\y \in C_M$ within one out of $2^d$ subsets depending on the signs of the variables $y_i$, $i=1, 2, ..., d$, and, for each subset, redefine the function of interest in such a way that we only have non-negative variables. To be specific, we can construct
\begin{eqnarray}
\breve \sff_{t,0}^M(y_{1:d}) &\dfn& \sff_t^M(y_{1:d}), \nonumber\\%&\mbox{ for }& y_1\ge 0, \ldots, y_d \ge 0, \nonumber\\
\breve \sff_{t,1}^M(y_{1:d}) &\dfn& \sff_t^M(y_{1:d-1},-y_d), \nonumber\\%&\mbox{ for }& y_1\ge 0, \ldots, y_{d-1} \ge 0, y_d<0, \nonumber\\
\breve \sff_{t,2}^M(y_{1:d}) &\dfn& \sff_t^M(y_{1:d-2},-y_{d-1},y_d), \nonumber\\%&\mbox{ for }& y_1\ge 0, \ldots, y_{d-2} \ge 0, y_{d-1}<0  y_d \ge 0\nonumber
&\vdots&\nonumber\\
\breve \sff_{t,2^d-1}^M(y_{1:d}) &\dfn& \sff_t^M(-y_{1:d}),
\end{eqnarray}
where we invert the sign of those variables $y_i < 0$. For each function $\breve \sff_{t,k}^M(\y)$, $k=0, 1, \ldots, 2^d-1$, we can repeat the analysis (over $\y \in [0, b_M]^d$) and arrive at the bounds 
\begin{equation}
\sup_{\y \in \left[ 0, \frac{1}{2}M^{\frac{\beta}{dp}} \right]^d} | \breve \sff_{t,k}^M(\y) | \le \frac{ V_{0,k}^\varepsilon }{ M^{\frac{1}{2}-\varepsilon} },
\quad k=0, 1, \ldots, 2^d-1, \nonumber
\end{equation} 
where $\varepsilon \in (0,\frac{1}{2})$ and every r.v. $V_{0,k}^\varepsilon$ is a.s. finite. Adding together the $2^d$ bounds, we obtain the inequality in \eqref{eqFtStatement}, with $V_t^\varepsilon = \sum_{k=0}^{2^d-1} V_{0,k}^\varepsilon$ an a.s. finite r.v., and conclude the proof.  \hfill \qed
%%%

Before stating the next partial result, let us recall assumption \mfC~again, namely the inequality $\mu_t(\overline{C_M})\le b M^{-\eta}$, where $b>0$ and $\eta<1$ are constants w.r.t $M$ and $\overline{C_M}$ is the complement of $C_M$.
%%%
\begin{Lema} \label{lmIAE} \label{lmIAE}
Let the sequence $\y_{0:T}$, $T<\infty$, be arbitrary but fixed and assume that \mfL, \mfD~and \mfC~hold. Then, for any $0 < \varepsilon < \frac{1}{2}$ and each $t=1, 2, ..., T$ there exists an a.s. finite r.v. $W_t^\varepsilon$ independent of $M$ such that
\begin{eqnarray}
\int | p_t^M(\y) - p_t(\y) | d\y \leq \frac{\tilde W_t^\varepsilon}{M^{\left(
	\frac{1}{2}-\varepsilon
\right) \wedge \eta}}.
\end{eqnarray}
\end{Lema}
%%%
\noindent\textbf{\textit{Proof}}:
We start with a trivial decomposition of the integrated absolute error,
\begin{eqnarray}
\int \left|
        p_t^M(\y) - p_t(\y)
\right| d\y &=& \int_{C_M} \left|
        p_t^M(\y) - p_t(\y)
\right| d\y \nonumber\\
&&+ \int_{\overline{C_M}} \left|
        p_t^M(\y) - p_t(\y)
\right| d\y \nonumber\\
&\le&  \int_{C_M} \left|
        p_t^M(\y) - p_t(\y)
\right| d\y \nonumber\\
&&+ 2\int_{\overline{C_M}} p_t(\y)d\y \nonumber \\
&& + \int_{\overline{C_M}} \left(
        p_t^M(\y) - p_t(\y)
\right) d\y, \label{eqGG0.5}
\end{eqnarray}
where the equality follows from $C_M \cup \overline{C_M} = \Real^{d_y}$ and the inequality is obtained from the fact that $p_t$ and $p_t^M$ are non-negative, hence $|p_t^M(\y)-p_t(\y)| \le p_t^M(\y) + p_t(\y)$. Moreover, if we realize that
\begin{eqnarray}
\int_{\overline{C_M}} \left(p_t^M(\y) - p_t(\y)\right)d\y &=& 1 - \int_{C_M} p_t^M(\y) d\y \nonumber \\
&& -1 + \int_{C_M} p_t(\y)d\y \nonumber\\
&=& \int_{C_M} \left( p_t(\y) - p_t^M(\y) \right)d\y\nonumber
\end{eqnarray}
then it is straightforward to see that
\begin{equation}
\int_{\overline{C_M}} \left(p_t^M(\y) - p_t(\y)\right)d\y \le \int_{C_M} \left|
        p_t^M(\y) - p_t(\y)
\right| d\y
\label{eqGG1}
\end{equation}
and, as a consequence, substituting \eqref{eqGG1} into \eqref{eqGG0.5},
\begin{eqnarray}
\int \left|p_t^M(\y) - p_t(\y)\right| d\y &\le&  
2\int_{C_M} \left|
        p_t^M(\y) - p_t(\y)
\right| d\y \nonumber\\
&&+ 2\int_{\overline{C_M}} p_t(\y)d\y \label{eqThreeTerms}
\end{eqnarray}
The first term on the right-hand side of \eqref{eqThreeTerms} can be bounded easily because $C_M$ is compact, namely 
 \begin{equation}
\int_{C_M} \left|
        p_t^M(\y) - p_t(\y)
\right| dx \le \mL(\overline{C_M}) \sup_{\y \in C_M} \left|
        p_t^M(\y) - p_t(\y)
\right|,
\label{eqIneq1}
\end{equation}
where $\mL(C_M) = ( 2b_M )^{d_y} = M^\frac{\beta}{p}$ is the Lebesgue measure of $C_M$. As for the supremum in \eqref{eqIneq1}, we only need to realise that the function $\sff_t^M(\y)=p_t^M(\y)-p_t(\y)=(g_t^\y,\xi_t^M) - (g_t^\y,\xi_t)$ belongs to the class $\sfF_t^M(d_y)$ under assumptions \mfL~and \mfD. Therefore, we can apply Lemma \ref{lmUnif} to show that $\sup_{\y \in C_M} | p_t^M(\y) - p_t(\y) | \le V_t^{\varepsilon_1}/M^{\frac{1}{2}-\varepsilon_1}$, where $V_t^{\varepsilon_1} \ge 0$ is an a.s. finite r.v.  and $\frac{1+\beta}{p} < \varepsilon_1 < \frac{1}{2}$ is a constant, both independent of $M$. Therefore, the inequality \eqref{eqIneq1} can be extended to yield
\begin{equation}
\int_{C_M} \left|
        p_t^M(\y) - p_t(\y)
\right| d\y \le  \frac{
        V_t^{\varepsilon_1}
}{
        M^{\frac{1}{2}-\varepsilon_1-\frac{\beta}{p}}
} =  \frac{
        V_t^{\varepsilon}
}{
        M^{\frac{1}{2}-\varepsilon}
},
\label{eqBoundFirstTerm}
\end{equation}
where $\varepsilon=\varepsilon_1 + \frac{\beta}{p}$ and $V_t^{\varepsilon}=V_t^{\varepsilon_1}$. If we choose $\varepsilon_1 < \frac{1}{2}-\frac{\beta}{p}$, then $\varepsilon \in \left( \frac{1+2\beta}{p}, \frac{1}{2} \right)$. Note that, for $\beta < 1$ and choosing $p \ge 6$, $\frac{1}{2}-\frac{\beta}{p}-\frac{1+\beta}{p} > \frac{1}{2}-\frac{3}{p} > 0$, hence both $\varepsilon_1$ and $\varepsilon$ are well defined. Now, taking $p$ large enough we can effectively select $\varepsilon \in (0,\frac{1}{2})$.

For the second integral in Eq. \eqref{eqThreeTerms}, note that $\int_{\overline{C_M}}p_t(\y)d\y = \mu_t(\overline{C_M})$ and, therefore, it can be bounded directly from assumption \mfC, i.e., 
\begin{equation}
2\int_{\overline{C_M}} p_t(\y)d\y \le 2 b M^{-\eta},
\label{eqBoundSecondTerm}
\end{equation}
where $b>0$ and $\eta>0$ are constant w.r.t. $M$.
Putting together Eqs. \eqref{eqThreeTerms}, \eqref{eqBoundFirstTerm}  and \eqref{eqBoundSecondTerm} yields the desired result, with $\tilde W_t^\varepsilon = 2(V_t^\varepsilon + b) < \infty$ a.s.
\hfill\qed
%%%

Finally, the proof of Theorem \ref{th1} is a straightforward application of Lemma \ref{lmIAE}.

%%%
\noindent\textit{\textbf{Proof of Theorem 1.}}
We first note that, for any bounded function $h$,
\begin{eqnarray}
\left( h, \mu_t^M \right) - \left( h, \mu_t \right) &=& \int h(y)p_t^M(y)dy - \int h(y)p_t(y)dy \nonumber\\
&=& \int h(y) \left( p_t^M(y) - p_t(y) \right)dy, \nonumber
\end{eqnarray}
hence, trivially,
\begin{equation}
\left| \left( h, \mu_t^M \right) - \left( h, \mu_t \right) \right| \le \|h\|_\infty \int \left| p_t^M(y) - p_t(y) \right|dy.
\label{eqkk9}
\end{equation}
If we apply Lemma \ref{lmIAE} on the right hand side of \eqref{eqkk9} then we readily obtain
\begin{equation}
\left| \left( h, \mu_t^M \right) - \left( h, \mu_t \right) \right| \leq \|h\|_\infty \frac{\tilde W_t^\epsilon}{M^{(\frac{1}{2}-\epsilon) \wedge \eta}},
\label{eqGG3}
\end{equation}
where $\epsilon \in (0,\frac{1}{2})$ is an arbitrarily small constant independent of $M$ and $W_t^\epsilon=\| h \|_\infty \tilde W_t^\varepsilon$ is an a.s. finite r.v., also independent of $M$.
\hfill\qed

%%%%%%%%%%%%%%%%%%%%%%
%%%%%%%%%%%%%%%%%%%%%%
%%%%%%%%%%%%%%%%%%%%%%
\section{Proof of Proposition  \ref{prop_indep}}
\label{appendix_prop_pmf}
%%%%%%%%%%%%%%%%%%%%%%
%%%%%%%%%%%%%%%%%%%%%%
%%%%%%%%%%%%%%%%%%%%%%

The sequence of r.v.'s $\{ A_{K,t} \}_{t \ge 1}$ are {\em constructed} to be independent. To see this, let us look into the generation of $A_{K,t}$ and $A_{K,t+1}$. Below, we are using capital letters to denote a r.v. (e.g., $Y_t$) and lower-case letters for its realisations (e.g., $y_t$). 

At time $t$, the r.v. $A_{K,t}$ is constructed by means of a nonlinear transformation of the r.v.'s $Y_t$ and $\{ \tilde Y_t^{(k)} \}_{k=1, ..., K}$. The latter are referred to as fictitious observations in the paper. Let {us} denote this many-to-one transformation as $\psi$, i.e., 
\begin{equation}
A_{K,t} = \psi(Y_t, \tilde Y_t^{(1)}, \ldots, \tilde Y_t^{(K)}).
\label{eqTransf}
\end{equation}
Under the sole assumption that $\{ Y_t, \tilde Y_t^{(1)}, \ldots, \tilde Y_t^{(K)} \}$ are i.i.d. continuous r.v.'s, Proposition \ref{prop_pmf} states that $A_{K,t}$ has a uniform probability distribution. To be precise, $A_{K,t}$ takes values on $\{ 0, ..., K \}$, and its probability mass function is $P(A_{K,t} = n) = \frac{1}{K+1}$ for every $n \in \{ 0, ..., K \}$.

In our case, the common pdf of the r.v.'s $\{ Y_t, \tilde Y_t^{(1)},$ $ \ldots, \tilde Y_t^{(K)} \}$ is $p_t(y_t) $ $ = \int g_t(y_t,\x) \xi_t(d\x) = p(y_t|y_{1:t-1})$. However, the actual form of $p(y_t|y_{1:t-1})$ plays no role whatsoever in Proposition \ref{prop_pmf}. In other words, $A_{K,t}$ is uniform as long as $\{ Y_t, \tilde Y_t^{(1)}, \ldots, \tilde Y_t^{(K)} \}$ are i.i.d. and this results holds independently of the actual sequence $y_{1:t-1}$ (which determines the form of $p(y_t|y_{1:t-1})$). 

We move on to time $t+1$. The r.v. $A_{K,t+1}$ is obtained as a nonlinear transformation of $\{ Y_{t+1}, \tilde Y_{t+1}^{(1)}, \ldots, \tilde Y_{t+1}^{(K)} \}$, namely,
$$
A_{K,t+1} = \psi(Y_{t+1}, \tilde Y_{t+1}^{(1)}, \ldots, \tilde Y_{t+1}^{(K)}).
$$ 
From Proposition 1, if $\{ Y_{t+1}, \tilde Y_{t+1}^{(1)}, \ldots, \tilde Y_{t+1}^{(K)} \}$ are i.i.d. then $A_{K,t+1}$ has a uniform distribution, i.e., $P(A_{K,t+1} = n) = \frac{1}{K+1}$ for every $n \in \{ 0, ..., K \}$. As before, this is true independently of the specific common pdf of the r.v.'s  $\{ Y_{t+1}, \tilde Y_{t+1}^{(1)}, \ldots, \tilde Y_{t+1}^{(K)} \}$. This common pdf is $p_{t+1}(y_{t+1})=(g^{y_{t+1}}_{t+1},\xi_{t+1})=p(y_{t+1}|y_{1:t})$ and, therefore, $A_{K,t+1}$ is uniform {without regard to} the sequence $y_{1:t}$ (which determines the form of $p(y_{t+1}|y_{1:t})$) and, in particular, {without regard to}  the observed realisation $Y_t=y_t$.

Now, since $A_{K,t+1}$ is uniform for any $Y_t=y_t$ (and, obviously, for any $\tilde Y_t^{(k)}=\tilde y_t^{(k)}$, $k=1, ..., K$), and $A_{K,t}$ is obtained as a transformation of $\{ Y_t, \tilde Y_t^{(1)}, \ldots, \tilde Y_t^{(K)} \}$ (see \eqref{eqTransf} above), then it follows that $A_{K,t+1}$ has a uniform distribution for every possible realisation $A_{K,t}=n$. This implies that the conditional distribution of $A_{K,t+1}$ given $A_{K,t}$ is uniform, i.e.,
\begin{eqnarray}
P(A_{K,t+1}=n | A_{K,t}=m ) =  \frac{1}{K+1},
\label{eqXasta}
\end{eqnarray}
$\forall (n,m) \in \{0, \ldots, K\} \times \{0, \ldots, K\}$. However, Eq. \eqref{eqXasta} readily entails independence. If we let $P(A_{K,t+1},A_{K,t})$ denote the joint probability mass function of $A_{K,t+1}$ and $A_{K,t}$, then from the definition of conditional probability we have
\begin{eqnarray}
P(A_{K,t+1}=n, A_{K,t}=m) &=& \nonumber\\
 P( A_{K,t+1}=n | A_{K,t}=m ) P( A_{K,t}=m )  &=&  \nonumber\\
 \frac{1}{K+1} \times \frac{1}{K+1}  &=&  \nonumber\\
P( A_{K,t+1}=n ) P( A_{K,t}=m ),
\end{eqnarray}
for any $n$ and $m$ within the set $\{ 0, \ldots, K \}$.

%%%%%%%%%%%%%%%%%%%%%%
%%%%%%%%%%%%%%%%%%%%%%
%%%%%%%%%%%%%%%%%%%%%%
\section{Proof of Theorem \ref{thQM}}
\label{apThQM}
%%%%%%%%%%%%%%%%%%%%%%
%%%%%%%%%%%%%%%%%%%%%%
%%%%%%%%%%%%%%%%%%%%%%

%%%
% Proof of convergence of \mbbQ_{K,M,t}
%%%
Let $Y_t$ denote the (random) observation at time $t$. Assume, without loss of generality, that $\mY = \Real$. The probability measure associated to $Y_t|Y_{1:t-1}=y_{1:t-1}$ is $\mu_t(dy)$ and, therefore, we can write the cumulative distribution function of  $Y_t|Y_{1:t-1}=y_{1:t-1}$ as $F_t(z) = (I_{(-\infty,z]},\mu_t)$, where
$$
I_A(y) = \left\{
	\begin{array}{ll}
	1, &\mbox{if $y \in A$}\\
	0, &\mbox{otherwise}\\
	\end{array}
\right.
$$
is the indicator function. Obviously, $\| I_A \|_\infty = 1 < \infty$ independently of the set $A$ and, therefore, Theorem \ref{th1} yields
$$
\lim_{M\rw\infty} F^M_t(z)=F_t(z) \quad \mbox{a.s.}
$$
for any $z\in \Real$, where $F^M_t(z)= (I_{(-\infty,z]},\mu^M_t)$ is the approximation of the cdf of $Y_t|Y_{1:t-1}=y_{1:t-1}$ provided by the BPF. 

Assume the actual observation is $Y_t=y_t$ and we draw $K$ i.i.d. fictitious observations $\tilde y_t^{(1)}, \ldots, \tilde y_t^{(K)}$ from the distribution with cdf $F_t^M$. Given $Y_t=y_t$ is fixed, the probability that exactly $n$ out of $K$ of these samples are lesser than $y_t$ coincides with the probability to have $n$ successes out of $K$ trials for a binomial r.v. with parameter (i.e., success probability) $F_t^M(y_t)$, which can be written as
$$
h_n^M(y_t) = {K \choose n} \left( F_t^M (y_t) \right)^n \left( 1 - F_t^M(y_t) \right)^{K-n}.
$$
By integrating $h_n^M(y_t)$ over the predictive distribution of $Y_t$, we obtain the probability to have exactly $n$ fictitious observations, out of $K$, which are less than the r.v. $Y_t$, i.e., the probability that $A_{K,M,t}=n$ is
\begin{equation}
\mathbb{Q}_{K,M,t}(n) = ( h_n^M,\mu_t).
\label{eqGG10}
\end{equation}
However, Theorem \ref{th1} yields $\lim_{M\rw\infty} (h_n^M,\mu_t^M) = (h_n^M,\mu_t)$ a.s.\footnote{Note that $\| h_n^M \|_\infty=1$ independently of $n$ and $M$. If we recall the proof of Theorem \ref{th1}, namely  inequality \eqref{eqGG3}, we observe that the error rates for the approximation errors of the form $| (h,\mu_t^M) - (h,\mu_t) |$ depend on the test function $h$ only through its supremum $\| h \|_\infty$, i.e., the r.v. $\tilde W_t^\varepsilon$ in \eqref{eqGG3} only depends on the observations $\y_{1:t-1}$ and the model (specifically the likelihood functions). Therefore, Theorem 1 (the same as, e.g., Lemmas \ref{lmXi-1} and \ref{lmPt-1}) also holds for any test function that depends on $M$ (even a random one) as long as its supremum is deterministic and independent of $M$. This is the case of function $h_n^M(y)$.} and, in particular, there exists a sequence of non-negative r.v.'s $\{ \varepsilon_M \}_{M\ge 1}$ such that $\lim_{M\rw\infty} \varepsilon_M=0$ a.s. and
\begin{equation}
(h_n^M, \mu_t^M) - \varepsilon_M \leq (h_n^M,\mu_t) \leq (h_n^M, \mu_t^M) + \varepsilon_M
\label{eqGG11}
\end{equation}
for each $M$. Moreover, it is apparent that $(h_n^M, \mu_t^M) = \frac{1}{K+1}$ (see Proposition \ref{prop_pmf}) which, together with \eqref{eqGG10} and \eqref{eqGG11} yields the desired relationship
$$
\frac{1}{K+1} - \varepsilon_M \le \mbbQ_{K,M,t}(n) \le \frac{1}{K+1}+\varepsilon_M
$$
for every $n \in \{0, ..., K \}$. \hfill \qed

%%%

%%%%%%%%%%%%%%%%%%%%%%
%%%%%%%%%%%%%%%%%%%%%%
%%%%%%%%%%%%%%%%%%%%%%
\bibliographystyle{IEEEbib}
%\bibliography{bibliografia}

\begin{IEEEbiography}[{\includegraphics[width=1in,height=1.25in,clip,keepaspectratio]{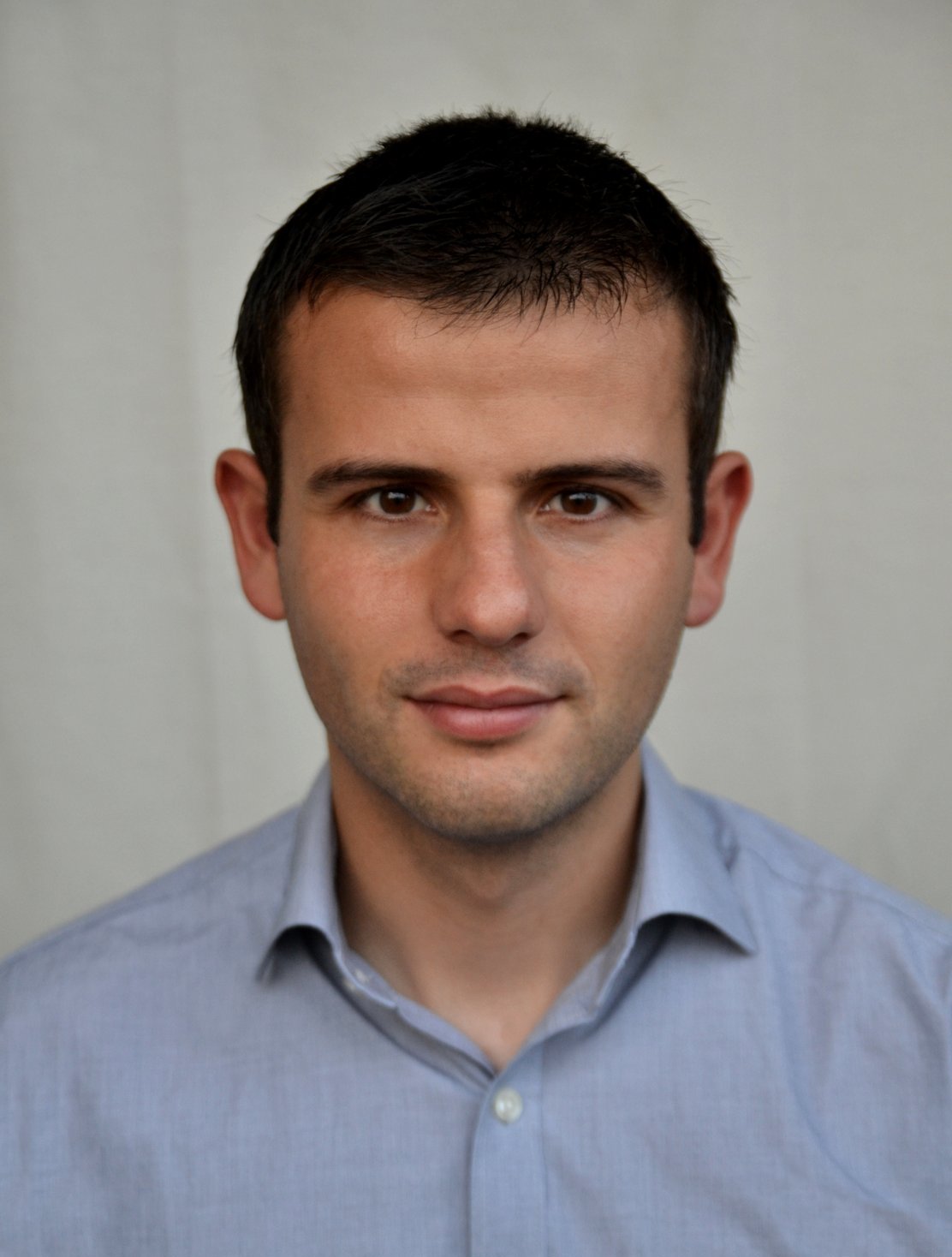}}]{V\'ictor Elvira}
 received the M.Sc. and Ph.D. degrees in electrical engineering from Universidad de Cantabria (Spain) in 2008 and 2011, respectively. In 2012, he joined Universidad Carlos III de Madrid (Spain) as postdoctoral researcher, and later as an Assistant Professor. In 2016, he joined IMT Lille Douai, an engineering school of the \emph{Institut Mines-T\'el\'ecom}, where he is currently an Associate Professor. He also belongs to the CRIStAL laboratory (UMR CNRS 9189). He has been a visiting scholar at the IHP Leibniz Institute (Frankfurt Oder, Germany), University of Helsinki (Finland),  Stony Brook University of New York (USA), Federal University of Rio de Janeiro (Brazil), and Paris-Dauphine University (France). His research interests include computational statistics, statistical signal processing, Bayesian analysis, and biomedical signal processing. He has co-authored over 40 journal and peer-reviewed conference papers. 
\end{IEEEbiography}
\begin{IEEEbiography}[{\includegraphics[width=1in,height=1.25in,clip,keepaspectratio]{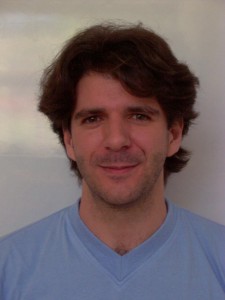}}]{Joaqu\'in M\'iguez}
received the M.Sc. and Ph.D. degrees in computer engineering from the University of A Coru\~na (A Coru\~na, Spain) in 1997 and 2000, respectively. He has held permanent positions at the Department of Electronics and Systems, University of A Coru\~na (2000--03), the School of Mathematical Sciences, Queen Mary University of London (2015--2016), and the Department of Signal Theory \& Communications, Universidad Carlos III de Madrid (2004--15 and 2016--present). He has also held visiting positions in the Department of Electrical \& Computer Engineering of the State University of New York at Stony Brook (2001) and the Department of Mathematics of Imperial College London (2013--14). His research interests are in the fields of applied probability, statistical signal processing, Bayesian analysis, dynamical systems and the theory and applications of Monte Carlo methods. Dr. M\'{\i}guez has co-authored over 50 international journal papers in the fields of signal processing, communications, mathematical physics, probability and statistics. He has delivered lectures and seminars on various European universities and research centres. He is a co-recipient of the IEEE Signal Processing Magazine Best Paper Award 2007.
\end{IEEEbiography}
\begin{IEEEbiography}[{\includegraphics[width=1in,height=1.25in,clip,keepaspectratio]{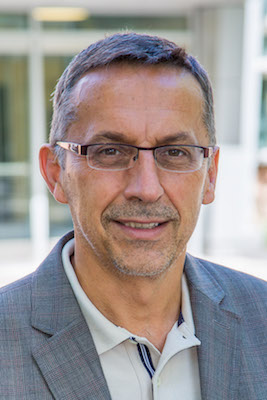}}]{Petar M. Djuri\'c} (M'90--SM'99--F'06) received the B.S. and M.S. degrees in electrical engineering from the University of Belgrade, Belgrade, in
1981 and 1986, respectively, and the Ph.D. degree in electrical engineering from the University of Rhode Island, Kingston, RI, in 1990. Since 1990, he has been a Professor with the Department of Electrical and Computer Engineering, Stony Brook University, Stony Brook, NY. From 1981 to 1986, he was a Research Associate with the Vin\v{c}a Institute of Nuclear Sciences, Belgrade. His research interests include the area of signal and information processing with primary interests in the theory of signal modeling, detection, and estimation; Monte Carlo-based methods; signal and information processing over networks; and applications in
a wide range of disciplines. He has been invited to lecture at many universities in the United States and overseas. He received the IEEE Signal Processing Magazine Best Paper Award in 2007 and the EURASIP Technical Achievement Award in 2012. In 2008, he was the Chair of Excellence of Universidad Carlos III de Madrid-Banco de Santander. From 2008 to 2009, he was a Distinguished
Lecturer of the IEEE Signal Processing Society. He has been on numerous committees of the IEEE Signal Processing Society and of many professional conferences and workshops. He is a Fellow of EURASIP and the Editor-in-
Chief of the IEEE Transactions on Signal and Information Processing over Networks.
\end{IEEEbiography}

% that's all folks
\end{document}